\documentclass[prd,eprint,showpacs,nofootinbib,amsfonts,amssymb,amsmath,notitlepage]{revtex4-1}
\usepackage{hyperref}

\usepackage{graphicx}
\usepackage{simplewick}
\usepackage{epstopdf}
\usepackage{multirow}
\usepackage{subfigure}
\usepackage{cases}
\usepackage{bm}
\usepackage{epsfig}
\usepackage{todonotes}
\usepackage{multirow}
\usepackage{units}
\usepackage[normalem]{ulem}

\def\be{\begin{equation}}
\def\ee{\end{equation}}
\def\bea{\begin{eqnarray}}
\def\eea{\end{eqnarray}}

\newcommand{\bear}{\begin{eqnarray}}

\newcommand{\eear}{\end{eqnarray}}

\newlength{\tskip}
\newbox\pippobox

\def\be{\begin{equation}}
\def\ee{\end{equation}}
\def\bea{\begin{eqnarray}}
\def\eea{\end{eqnarray}}

\def\9{\nabla}

\def\dd{{\rm d}}

\def\6{\partial}

\def\0{(0)}

\def\>{\rightarrow}

\begin{document}

\title{{\bf Searching for primordial localized features with CMB and LSS spectra}}

\author{ Bin Hu$^1$ and  Jes{\'u}s Torrado$^1$}
\affiliation{
\smallskip
$^{1}$ Institute Lorentz, Leiden University, PO Box 9506, Leiden 2300 RA, The Netherlands}

\begin{abstract}
Inspired by the study of mild transient reductions in the speed of sound of the adiabatic mode during inflation, we search for a primordial localized feature imprinted in cosmic microwave background and large-scale structure formation observables. We find some common oscillatory patterns both in the Planck CMB temperature-temperature power spectrum and the WiggleZ galaxy spectrum. By performing independent searches with these two data sets, we find a coincidence in the most significant mode previously found by Ach{\'u}carro \textit{et al.}\ 2013 by using only Planck data. Furthermore, the joint data analysis shows that the oscillation frequency of the feature gets better constrained, and the amplitude marginally deviates from zero, unlike what was observed using only Planck data. Besides the parameter estimation, we also discuss the Bayesian evidence. The addition of WiggleZ data mildly enhances the significance of the best mode found in the Planck data.
\end{abstract}


\maketitle

%%%%%%%%%%%%%%%%%%%%%%%%%%%%%%%%%%%%%%%%%%%%%%%%%%%%%%%%%%%%%%%%%%%%%%%%%%%%%%%%%%%%%%%%%%%%%%%%%%%%%

%% cccccccccccccccccccccccccccccccccccccccccccccc
\section{Introduction}
There exist several hints of oscillatory signals in the cosmic microwave background (CMB) observables, such as power spectrum \cite{Bennett:2012zja,Ade:2013uln} and bispectrum \cite{Ade:2013ydc}.  This motivates a search for such kind of features produced by inflationary scenarios beyond canonical single-field.\footnote{By canonical single-field we mean slow-roll regime, Bunch-Davies vacuum and canonical kinetic terms, with $c_s=1$.} Several mechanisms that produce oscillatory features have been realized, such as a transient reduction in the speed of sound \cite{Achucarro:2012fd,Achucarro:2013cva,Achucarro:2014msa}, step inflaton potential \cite{Starobinsky:1992ts,Adams:2001vc,Gong:2005jr,Chen:2008wn,Arroja:2011yu,Martin:2011sn,Adshead:2011jq,Arroja:2012ae,Bartolo:2013exa,Cannone:2014qna,Covi:2006ci,Benetti:2011rp,Benetti:2013cja,Hamann:2007pa,Benetti:2012wu,Stewart:2001cd,Choe:2004zg,Dvorkin:2009ne,Adshead:2011bw,Miranda:2012rm,Adshead:2013zfa}, different initial vacuum states \cite{Danielsson:2002kx,Greene:2004np,Meerburg:2009ys,Jackson:2010cw}, multi-field dynamics \cite{Gao:2012uq,Gao:2013ota,Saito:2013aqa,Noumi:2013cfa, Chen:2014joa,Chen:2012ja}, or phenomenological superimposed oscillations in the primordial power spectrum \cite{Martin:2003sg,Flauger:2009ab,Flauger:2010ja,Aich:2011qv,Meerburg:2011gd,Peiris:2013opa,Meerburg:2013cla,Meerburg:2013dla,Meerburg:2014kna}. In this work we focus on searching for oscillatory features in the scenario of a transient reduction in the speed of sound. The effect of a variable speed of sound has also been analyzed both in the power spectrum \cite{Achucarro:2010da,Hu:2011vr,Achucarro:2012fd} and bispectrum \cite{Achucarro:2012fd,Adshead:2013zfa,Ribeiro:2012ar}. Similar studies of power spectrum \cite{Park:2012rh,Miranda:2012rm,Nakashima:2010sa,Bean:2008na,Bartolo:2013exa} and bispectrum \cite{Bean:2008na,Bartolo:2013exa,Meerburg:2010ks} have also been done for models with sudden variations. In addition, the Planck collaboration searched for features in the CMB bispectrum for a number of theoretically motivated templates \cite{Ade:2013ydc}, including oscillatory templates. Although in none of these cases the statistical significance of the extended models has been found to be high enough to claim a detection, with the improvement of experimental accuracy we are now at the threshold of verifying or falsifying these models. 

Our test case, introduced in \cite{Achucarro:2013cva}, consists of a gaussian reduction in the speed of sound occurring within the window of $e$-folds corresponding to the angular scales probed by CMB and large-scale structure (LSS) surveys. 
Its functional form is consistent with a reduction in the speed of sound resulting from a soft turn along the inflationary trajectory in a multi-field theory in which the mass hierarchy is large enough to allow for an effective single-field description \cite{Achucarro:2010jv,Achucarro:2010da,Cespedes:2012hu,Achucarro:2012sm} (though one should keep in mind that a similar reduction in the speed of sound may result from a different high-energy completion of the effective field theory).

Since it is the same curvature perturbations that set the initial conditions for CMB anisotropies and large-scale structure distributions,  the primordial oscillatory signals should be imprinted in all the observables of CMB anisotropy and LSS tracers, like CMB spectra, bispectra, galaxy spectra, etc. Based on this consideration, in this paper we search for primordial oscillatory features from a transient reduction in the speed of sound of adiabatic curvature perturbations via both CMB anisotropy temperature-temperature spectrum of the Planck satellite as well as galaxy distribution spectrum of the WiggleZ telescope. The rest of this paper is organized as follows. In Sec.~\ref{review}, we will briefly review the theoretical setup of the transient reductions in the speed of sound. In Sec.~\ref{method}, we will introduce the methodology of parameter estimation and model selection which is adopted in this work as well as the data sets used. Then, we arrive at our results and discuss them in Sec.~\ref{discussion}. Finally, we conclude in Sec.~\ref{conclusion}.

\section{\label{review}Review of the model}
In the framework of effective field theory, heavy fields coupled to the inflaton reduce the speed of sound of the adiabatic mode each time the background inflationary trajectory makes a turn. In reference \cite{Achucarro:2012fd}, Ach{\'u}carro \textit{et al.}\ prove how small but abrupt changes in the speed of sound of the adiabatic mode during inflation, independently of their physical origin, seed discriminable features in the primordial power spectrum and bispectrum.

Defining $u\equiv1-c_s^{-2}$, at first order in this quantity a small transient reduction in the speed of sound produces a feature in the primordial scalar curvature power spectrum of the form
\begin{equation}
\label{eq:deltappfourier}
\frac{\Delta \mathcal{P}_\mathcal{R}}{\mathcal{P}_\mathcal{R}}(k)=k\int_{-\infty}^{0}d\tau\ u(\tau)\sin{(2k\tau)}\ ,
\end{equation}
where $\mathcal{P}_\mathcal{R}=H^2/(8\pi^2\epsilon m_{\text{Pl}}^2)$ is the featureless power spectrum with constant speed of sound $c_s=1$, and $\tau$ is the conformal time. The corresponding feature in the primordial bispectrum can also be found in \cite{Achucarro:2012fd}, and its calculation is valid in the perturbative regime of small $s\equiv \dot{c}_s/(c_s H)$. In the whole framework, it is assumed that the effect of the reduction in the speed of sound in both spectrum and bispectrum dominates over slow roll contributions. Both perturbativity and dominance over slow roll effects set bounds for the shape and the size of speed of sound reduction as
\begin{equation}
  \label{eq:usbounds}
  \mathcal{O}(\epsilon,\eta) ~\ll~ u,|s| ~\ll~ 1
  \ . 
\end{equation}

In this paper, following references \cite{Achucarro:2013cva,Achucarro:2014msa}, we search for reductions in the speed of sound which take the form of a gaussian in $e$-folds $N$:
\be
u=1-c_s^{-2}=B\,e^{-\beta(N-N_0)^2}=B\,e^{-\beta\left(\ln\frac{\tau}{\tau_0}\right)^2}\ ,
\label{eq:gaussefolds}
\ee
where we have introduced the parameters of the feature: the amplitude $B<0$, the sharpness $\beta>0$, and the instant of maximal reduction $N_0$ (or $\tau_0<0$).\footnote{Assuming that the slow-roll regime remains uninterrupted, conformal time and the $e$-folds time scale are related by $\ln\left(-\tau\right)=\left(N_\text{in}-N\right)-\ln\left(a_\text{in}H_0\right)$, where $a_\text{in}=a(N_\text{in})$ is the value of the scale factor when there were $N_\mathrm{in}$ $e$-folds left until the end of inflation -- we take here $N_\mathrm{in}=60$.} Eq.\ \eqref{eq:usbounds} imposes limits on the maximum and minimum values of the first two parameters, $B$ and $\beta$, while $\tau_0$, from the theoretical point of view, lacks a lower bound. One can set more conservative bounds on those parameters by imposing that the feature falls within the observable window of inflation in the CMB $N_\mathrm{CMB}$, and it is at the same time sharp enough and oscillates with large enough a frequency not to be degenerated with the cosmological parameters. If $N_\text{CMB}\simeq 7$ are the first 7 $e$-folds of the last $\sim 60$ of inflation, the constaints take the form \cite{Achucarro:2014msa}
\begin{subequations}
\label{eq:bounds}
\begin{gather}
\mathcal{O}(\epsilon,\eta)  \ll |B|  \ll 1\ ,\label{eq:bound1}\\
\frac{50}{N_\text{CMB}^2}    <  \beta  \ll \frac{2e}{B^2}\ ,\label{eq:bound2}\\
\frac{5}{\sqrt{2\beta}}    <   N_0-N_\text{in} < N_\text{CMB}-\frac{5}{\sqrt{2\beta}}\label{eq:bound3}
\ .
\end{gather}
\end{subequations}
That observability constraint sets in particular the lower limit of $\beta$ and both limits of $N_0$, giving bounds which are more conservative than the theoretical ones. Finally, the actual range of $\tau_0$ is further reduced to lie in the range $4.3\le\ln\left(-\tau_0\right)\le6.0$, motivated by a search for oscillatory features in the primordial bispectrum \cite{Achucarro:2013cva}.

%% cccccccccccccccccccccccccccccccccccccccccccccc
\section{\label{method}Methodology and Data sets}

In this paper we solve the Einstein-Boltzmann hierarchy by using CAMB \cite{Lewis:1999bs} and sample the parameter space using different approaches in order to fulfil two different purposes. On one hand, for parameter estimation, we use the thermodynamic Markov chain Monte Carlo (MCMC) sampler, \textsc{CosmoMC} \cite{Lewis:2002ah}. In detail, we use a Metropolis-Hastings algorithm to generate chains of samples for a set of cosmological parameters. On the other hand, for Bayesian evidence computation and model selection, we adopt the multimodal nested sampler, \textsc{MultiNest} \cite{Feroz:2008xx,Feroz:2007kg,2013arXiv1306.2144F} which implements an extended form of the nested sampling algorithm \cite{Skilling,Sivia,Mukherjee:2005wg,Liddle:2007fy,Shaw:2007jj}. This is because the dependence of the evidence on the prior requires that the prior space is adequately sampled, even in the regions of low likelihood. This makes evidence evaluation at least an order of magnitude more costly than parameter estimation. 

In what follows we make a brief review of the concepts of evidence and Bayesian ratio.
The Bayesian ratio is defined as the ratio of the probabilities of each of the two models conditioned on a given set of data $\mathbf{D}$:
\begin{equation}
\label{BE2}
R=\frac{P(M_1|\mathbf{D})}{P(M_0|\mathbf{D})}
 =\frac{\mathcal{Z}_1}{\mathcal{Z}_0} \frac{P(M_1)}{P(M_0)}
 =\frac{\mathcal Z_1}{\mathcal Z_0}\;.
\end{equation}
Here, $P(M_1)/P(M_0)$ is the probability ratio for the two models {\it a priori}, which is conventionally set to unity; the \emph{evidence} $\mathcal{Z}$ of a model $M$ is the marginalized likelihood of the data, i.e.\ the probability of having obtained the data $\mathbf{D}$ integrated over all possible values of the model parameters $\boldsymbol\theta$:
\begin{equation}
\label{Z2}
\mathcal{Z}=\int \mathcal{L}(\mathbf{D}|M(\boldsymbol\theta))\,\pi({\boldsymbol\theta})\,\dd^D{\boldsymbol\theta}\;,
\end{equation}
where $\mathcal{L}(\mathbf{D}|M(\boldsymbol\theta))$, $\pi({\boldsymbol\theta})$ and $D$ are, respectively, the likelihood of the data, the prior of the parameters in the model and the dimensionality of the parameter space. In this work, we will use $M_1$ and $M_0$ to denote the feature and featureless $\Lambda\mathrm{CDM}$ models;\footnote{$\Lambda$CDM denotes the 6-parameter base model considered by the Planck collaboration \cite{Ade:2013zuv}.} the cosmological parameter ranges we studied are listed in Tab.\ref{tab:parameters}. And the multidimensional integration in Eq.\ \eqref{Z2} was sampled via the multi-modal implementation of the nested sampling algorithm \textsc{MultiNest} \cite{Feroz:2008xx,Feroz:2007kg,2013arXiv1306.2144F}.

%% *-*-*-*-*-*-*-*-*-*-*-*-*-*-*-*-*-*-*-*-*-*-*-*-*-*-*-*-*-*-*-*-*-*-*-*-*-*-*-*-*-*-*-*-*-*-*-*
\begin{table}[ht!]
%\begin{ruledtabular}
\begin{tabular}{c|c|c}
\hline\hline
Parameter & \multicolumn {2}{c}{Range (min, max)}  \\
\hline
$\Omega_b h^2$ & \multicolumn {2}{c}{$(0.005,
0.100)$} \\
$\Omega_c h^2$ &  \multicolumn {2}{c}{$(0.01, 0.99)$}  \\
$100\vartheta_*$ &
\multicolumn {2}{c}{$(0.5, 10.0)$ }  \\
$\tau_\mathrm{reio}$ & \multicolumn {2}{c}{$(0.01, 0.80)$ } \\
$n_s$ &  \multicolumn {2}{c}{$(0.9, 1.1)$} \\
$\ln (10^{10} A_s^2)$ &  \multicolumn {2}{c}{$(2.7, 4.0)$} \\
\hline\hline
$ B$  &  \multicolumn {2}{c}{$(-0.2,0)$} \\
$\ln\beta$   & \multicolumn {2}{c}{$(0,7.5)$} \\
$\ln(-\tau_0)$       & \multicolumn {2}{c}{$(4.3,6.0)$} \\
\hline\hline
\end{tabular}
%\end{ruledtabular}
\caption{\label{tab:parameters}
List of the parameters used in the multimodal nested sampling. Besides these parameters, we also sample and marginalise over the fourteen nuisance parameters of the Planck likelihood and one bias parameter of the WiggleZ likelihood. We have sampled $B$ up to $-0.5$, but nothing interesting was found beyond the upper value cited in this table.}
\end{table}

The Bayesian evidence, Eq.\ \eqref{Z2}, measures the predictivity of a model. The integral is bigger the more amount of likelihood mass falls inside regions with substantial prior probability. The evidence is penalised by the volume $\mathcal{V}$ of the parameter space allowed by the theory, since the prior density goes roughly like $\pi\sim \mathcal{V}^{-1}$. In turn, the Bayesian ratio quantifies the relative predictivity of two models given a data set: if its value is much smaller than one, the model $M_0$ is a more likely explanation of the data than the model $M_1$, and vice versa. In the frequentist approach, this is comparable to the increase of $p$-values\footnote{From \href{http://en.wikipedia.org/wiki/P-value}{Wikipedia.org}, ``a $p$-value is the probability of obtaining a test statistic result at least as extreme as the one that was actually observed, assuming that the null hypothesis is true. A researcher will often ``reject the null hypothesis" when the $p$-value turns out to be less than a predetermined significance level, often $0.05$ or $0.01$. Such a result indicates that the observed result would be highly unlikely under the null hypothesis''.} due to the \emph{look-elsewhere effect}. For example, in particle physics, if one allows the predicted mass of a particle to vary within a broad range, the $p$-value of an apparent peak in particle production with a corresponding mass within this range will increase, just because a wider range of energies makes a random, non-physical peak-like feature more likely. Correspondingly, this indicates that the evidence of this model with a new parameter, like the new particle's mass, gets reduced.

In the particular case of localized primordial features in the CMB and LSS spectra, the Bayesian approach is motivated by the similarity that said features share with shot noise in the corresponding bands. This similarity, when the features are small, will result in the multi-modality of the likelihood of the corresponding parameters, and likelihood enhancements similar to those obtained by fitting the model to feature-less, noisy data. For example, for a specific linear oscillation template, using 5000 Planck-like, signal-less simulated CMB maps, the authors of \cite{Meerburg:2014kna} found that the noise could account for up to $\Delta\chi^2\equiv 2\Delta\ln\mathcal{L}\sim 30$ at $3\sigma$ confidence level, with a typical enhancement of $\Delta\chi^2\sim 10$ for the best fit of this kind of model. Considering this, it is not easy to assess whether we are fitting noise based on the likelihood enhancement only. Therefore, we focus on the predictivity of the models, given by their Bayesian evidence, to decide on the presence of features in the data. As explained above, in order to derive a reliable value for the evidence, we adopt a muti-modal nested sampling method. We assume flat priors for all the cosmological and nuisance parameters.

%\textcolor{red}{Similar circumstances happened in the searching for the localized primordial features. Because we are targeting for a localized signals, which share lots of similarities with short noise in the corresponding bands, this will result in likelihood multimodality of the corresponding parameters. For example, for a specific linear oscillation template, via 5000 Planck-like null signal simulations the authors of \cite{Meerburg:2014kna} found that the noise could account for $\Delta\chi^2\sim 30$ at $3\sigma$ confidence level, with a typical $\Delta\chi^2\sim 10$ improvements from the best fit of this kind of models. Given this consideration, in order to justify/falsify that we are not fitting noise, we have to go for the evidence. As explained above, in order to derive a reliable evidence, we adopt the muti-modal nested sampling and .}

On the other hand, the Bayesian ratio can also be used as an indicator of the correlation between two data sets with respect to an extended model $M_1$ based on a simpler model $M_0$: if the predictivity of the extended model with respect to the basis model increases when adding the new data set, this is an indication of the regions of high probability in the likelihood of the extended model being similar in the two data sets. Otherwise, the product of the likelihoods of both data sets would amount to a smaller evidence ratio than that of the single data sets.

As for the data sets, we use the measurements of CMB temperature anisotropy\footnote{\url{http://pla.esac.esa.int/pla/aio/planckProducts.html}} \cite{Ade:2013zuv}
from the first data release of the Planck survey. Its temperature power-spectrum likelihood is divided into low-$\ell$ ($\ell<50$) and high-$\ell$ ($\ell\geq 50$) parts.\footnote{This is because the central limit theorem ensures that the distribution of CMB angular power spectrum $C_{\ell}$ in the high-$\ell$ regime can be well approximated by Gaussian statistics. However, for the low-$\ell$ part the $C_{\ell}$ distribution is non-Gaussian. For these reasons the Planck team adopts two different methodologies to build the likelihood. In detail, for the low-$\ell$ part, the likelihood exploits all Planck frequency channels from $30$ to $353$ GHz, separating the cosmological CMB signal from diffuse Galactic foregrounds through a physically motivated Bayesian component separation technique. For the high-$\ell$ part, a correlated Gaussian likelihood approximation is employed. This is based on a fine-grained set of angular cross-spectra derived from multiple detector power-spectrum combinations between the $100$, $143$, and $217$ GHz frequency channels, marginalizing over power-spectrum foreground templates.} In order to break the well-known parameter degeneracy between the reionization optical depth $\tau_\mathrm{reio}$ and the scalar index $n_s$, the low-$\ell$ WMAP polarization likelihood (WP) is used \cite{Ade:2013zuv}. Finally, the unresolved foregrounds are marginalized over, assuming wide priors on the relevant nuisance parameters as described in \cite{Ade:2013kta}. 

Since several interesting feature modes are hinted at by using only Planck temperature-temperature spectrum in the study of Ach{\'u}carro {\it et al.} \cite{Achucarro:2013cva}, a natural step is to cross check these results with other observables seeded by the same initial conditions, coming from different experiments whose systematic uncertainties are different from Planck's. We use the measurements of the galaxy power spectrum made by the WiggleZ Dark Energy Survey.\footnote{\url{http://smp.uq.edu.au/wigglez-data}} As described in \cite{Parkinson:2012vd}, we use the power spectrum measured from spectroscopic redshifts of $170\,352$ blue emission line galaxies over a volume of $\sim\unit[1]{Gpc^{3}}$ \cite{Drinkwater:2009sd}. The covariance matrices as given in \cite{Parkinson:2012vd} are computed using the method described by \cite{Blake:2010xz}. The best model proposed for non-linear corrections to the matter power spectrum was calibrated against simulations. 
The surveys scan seven fields. Three of them are in the northen hemisphere, $9$-hr, $11$-hr and $15$-hr; and four in the southern hemisphere, $22$-hr, $1$-hr, $3$-hr and $0$-hr regions. Furthermore, the resulting galaxy spectra are constructed in four redshift bins, namely $0.1<z<0.3$, $0.3<z<0.5$, $0.3<z<0.7$ as well as $0.7<z<0.9$.
The likelihood in each redshift bins assumes Gaussian form
\begin{equation}
-2\log\mathcal{L}=\sum_{i,j}\Delta P_i \mathbf{C}^{-1}_{ij}\Delta P_j\;\;, {\rm with}\;\; \Delta P_i = P^{\rm th, con}_i - P^{\rm obs,g}_i\;.
\end{equation}
$\mathbf{C}_{ij}$ is the covariance matrix of the galaxy power spectrum. $P^{\rm th, con}_i$ is the $i$-th wavenumber band of the theoretical galaxy power spectrum, $P_{\rm g}(k)=b^2P_{\rm m}(k)$, convolved with WiggleZ window function $\mathbf{W}_{ij}$
\begin{equation}
\label{convg}
P^{\rm th, con}_i(k)=\sum_j\frac{\mathbf{W}_{ij}(k)P^{\rm th, g}_{j}(k/a_{\rm scl})}{a^3_{\rm scl}}\;,
\end{equation}
and $a_{\rm scl}$ is the scaling, which takes into account the observed galaxy redshift-space positions are converted to real space position using a fiducial cosmology. 
 $b$ is the linear galaxy bias against matter power spectrum, $P_{\rm m}(k)$
\begin{equation}
b^2 = \frac{\sum_{j,k}P^{\rm th, con}_j\mathbf{C}^{-1}_{jk}P^{\rm obs,g}_k}{\sum_{j,k}P^{\rm th, con}_j\mathbf{C}^{-1}_{jk}P^{\rm th, con}_k}\;.
\end{equation}
In the following analysis, we analytically marginalise over a linear galaxy bias in each of the four redshift bins by using the above expression.
It has already been demonstrated that linear theory predictions are as good a fit to the data as the calibrated model up to $k \sim 0.2\,h/\unit{Mpc}$  \cite{Parkinson:2012vd,Riemer-Sorensen:2013jsa}. For these reasons we restrict ourselves to scales smaller than $k_\mathrm{max} = 0.2\,h/\unit{Mpc}$ and use the linear theory prediction only.

%% cccccccccccccccccccccccccccccccccccccccccccccc
\section{\label{discussion}Results and Discussion}
In order to justify or falsify this model, we should go beyond CMB observables from the Planck satellite. A feature in the primordial spectrum of density perturbations will seed both CMB anisotropies and the tracers of matter perturbation, such as the galaxy distribution. Thus, if those features are big enough we should observe them via all those windows. 

%%---- fig zoomin ----
\begin{figure*}[t!]
\centering
\subfigure[\label{Fig:15hr_red1}]{\includegraphics[width=0.47\textwidth]{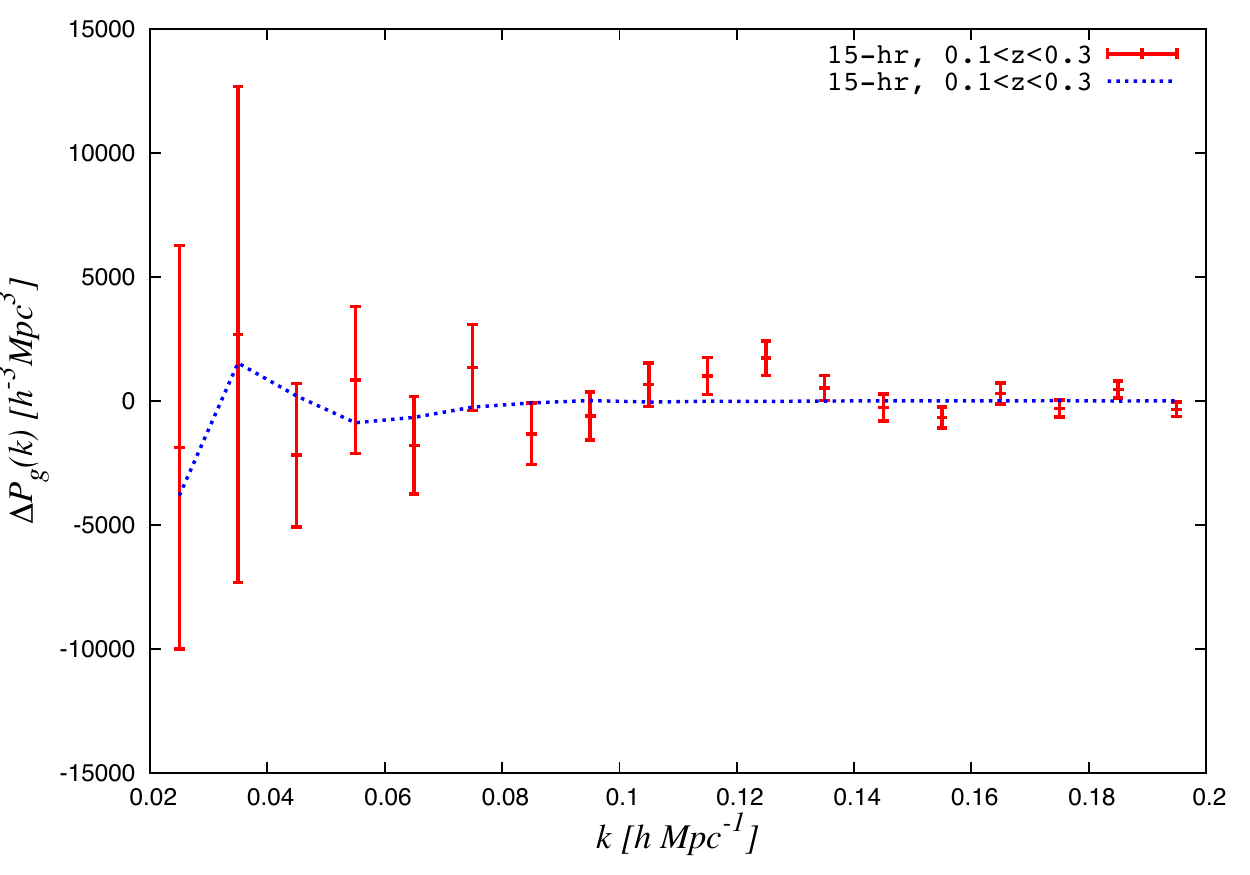}}
\subfigure[\label{Fig:15hr_red2}]{\includegraphics[width=0.47\textwidth]{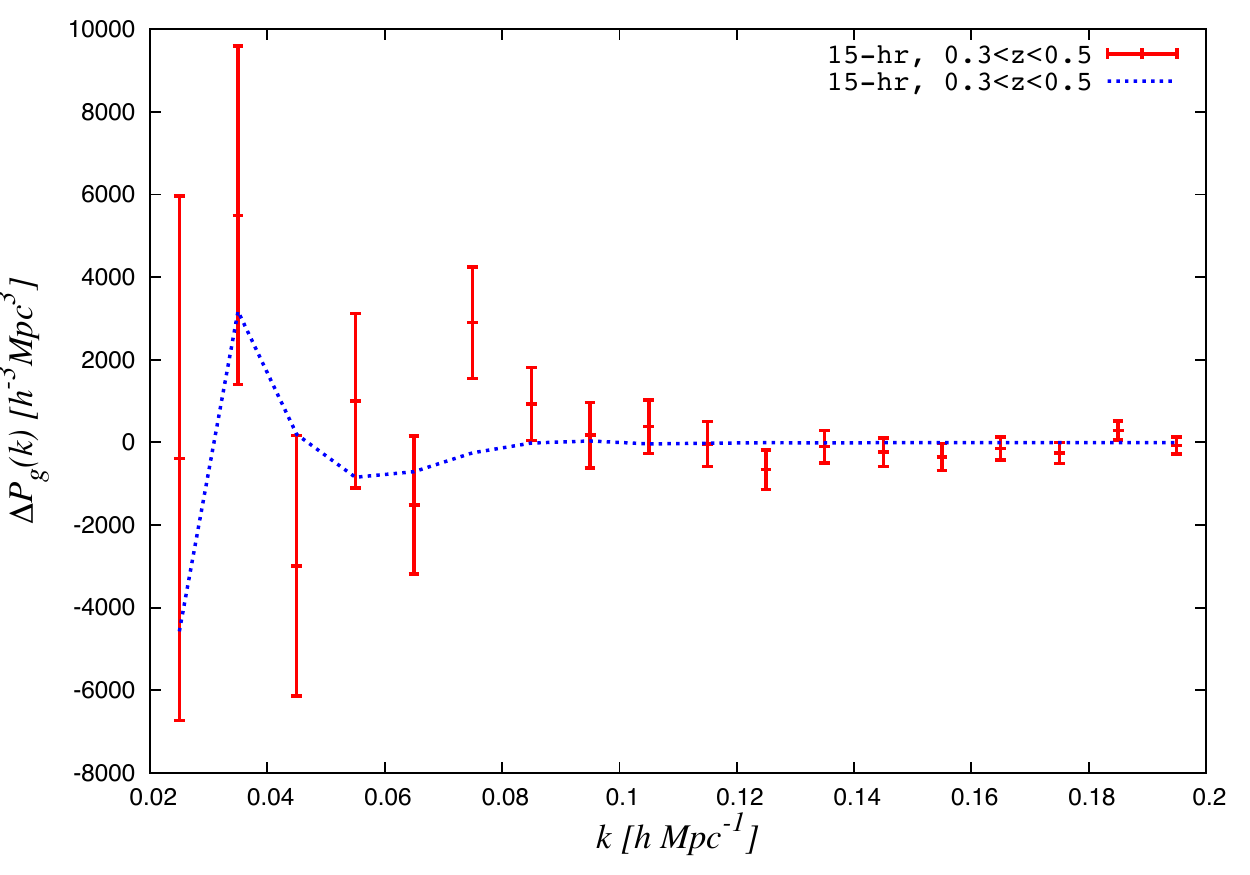}}
\subfigure[\label{Fig:15hr_red3}]{\includegraphics[width=0.47\textwidth]{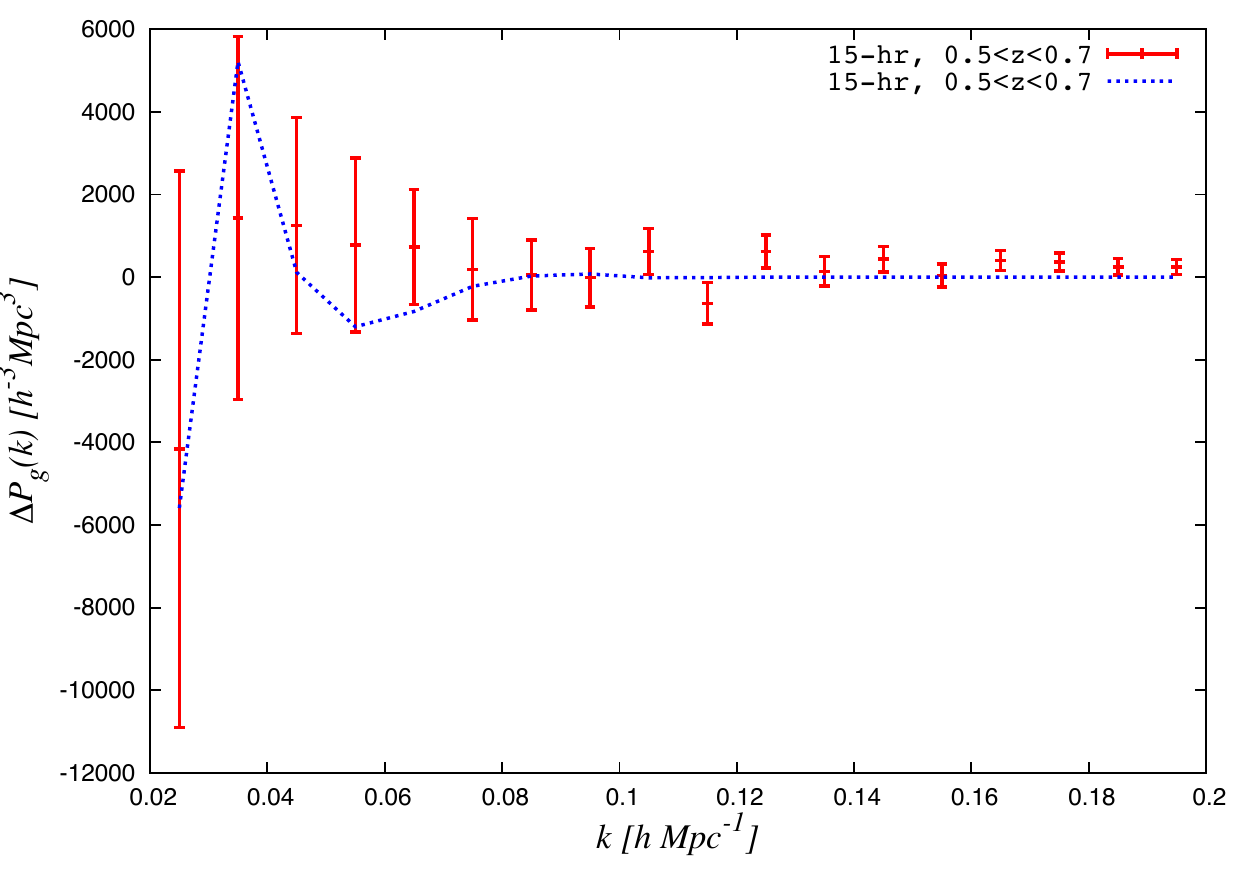}}
\subfigure[\label{Fig:15hr_red4}]{\includegraphics[width=0.47\textwidth]{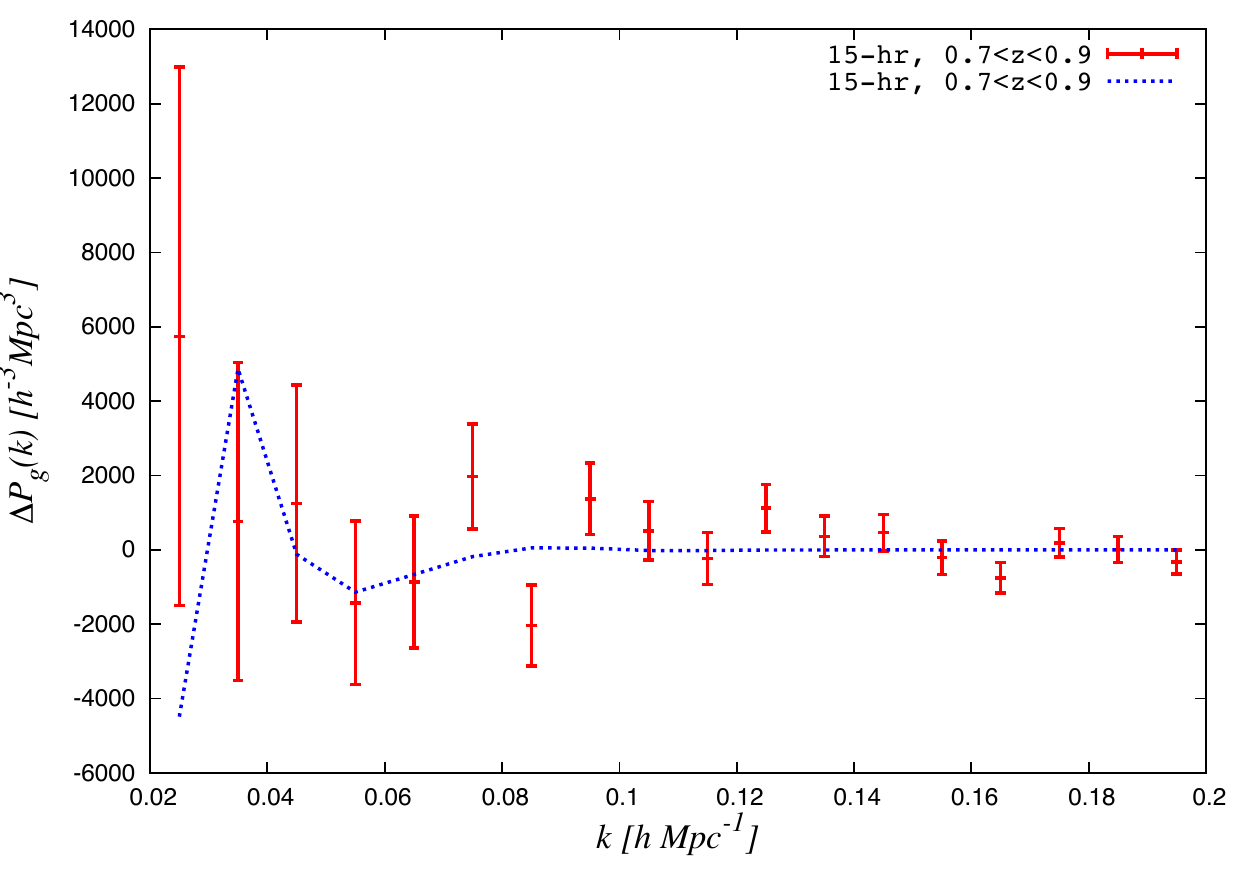}}
\caption{\label{Fig:15hr}
Galaxy power spectrum residuals of the feature model against the base $\Lambda$CDM model in four redshift bins. The data points come from the 15-hr region of WiggleZ survey. The curves are the differences in the convolved galaxy spectra, eq.\ \ref{convg}, between the feature model with parameter values $B=-0.045$, $\ln(-\tau_0)=5.55$, $\ln(\beta)=6.3$, and the base $\Lambda$CDM model.}
\end{figure*}

\begin{figure*}[t!]
\centering
\subfigure[WiggleZ\label{Fig:ZoomIn:WiggleZ}]{\includegraphics[width=0.47\textwidth]{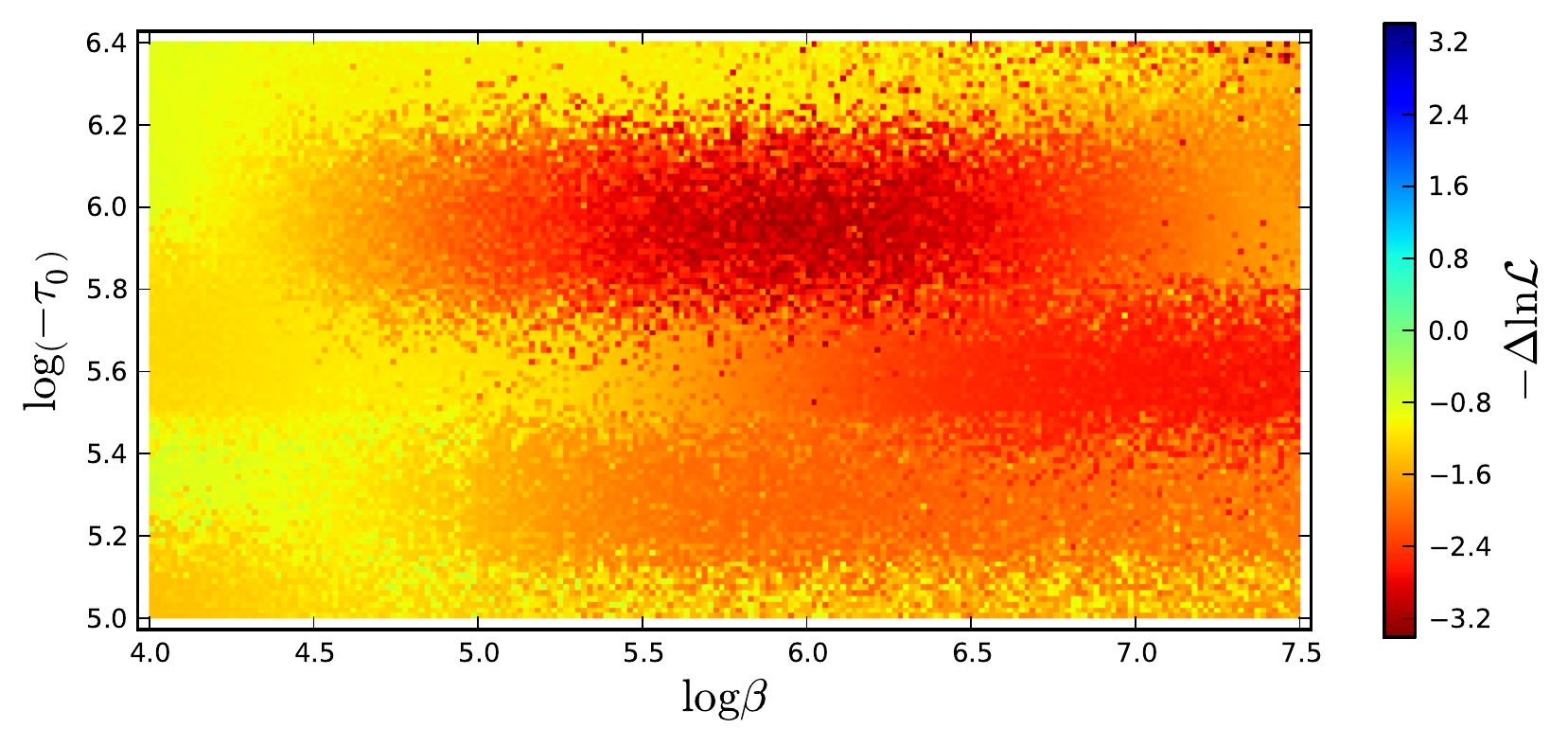}}
\subfigure[Planck\label{Fig:ZoomIn:Planck}]{\includegraphics[width=0.47\textwidth]{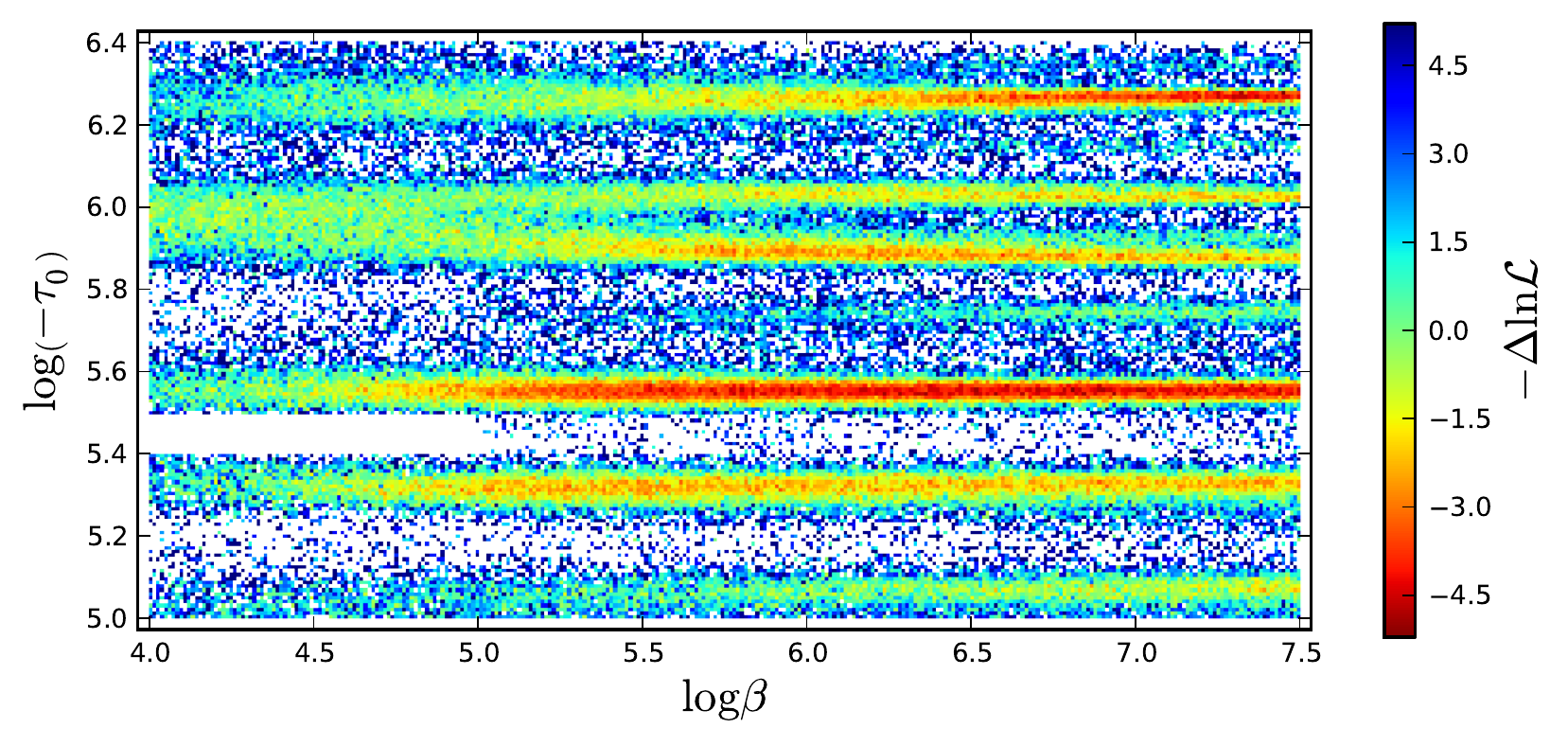}}
\subfigure[Planck$+$WiggleZ\label{Fig:ZoomIn:PlanckWiggleZ}]{\includegraphics[width=0.6\textwidth]{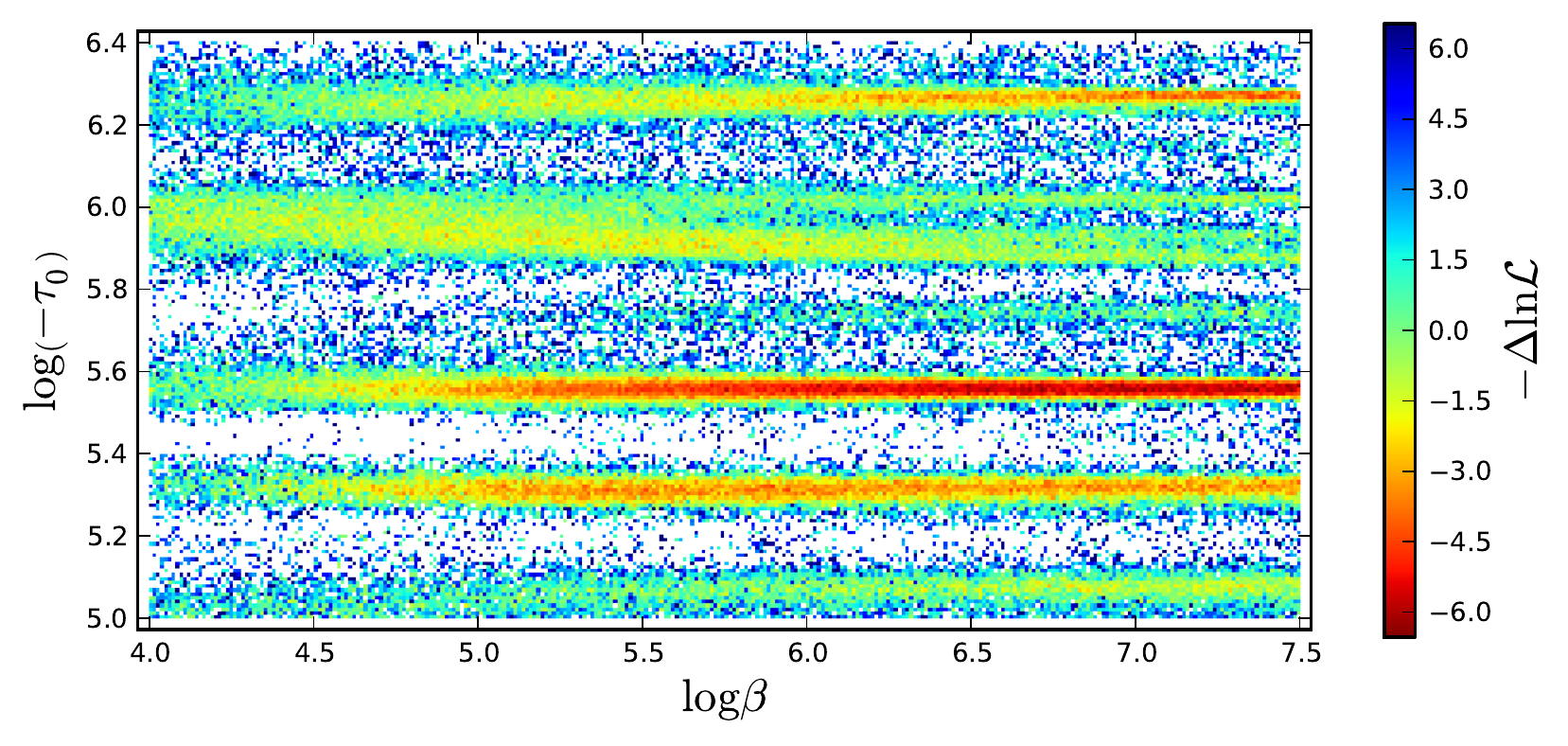}}
\caption{\label{Fig:ZoomIn}
Profile parameter distribution of the MCMC sampling in the $(\ln\beta,-\ln(-\tau_0))$ plane, for the different combinations of data sets. It shows the coincidence between the fits found in Planck and WiggleZ at $\ln(-\tau_0)\sim 5.3$ and $\ln(-\tau_0)\sim 5.55$, and their enhancement of 20\% in likelihood improvement. The difference in the likelihood ($\Delta$) is calculated against the best fit value of $\Lambda\mathrm{CDM}$ in the different data sets. The regions where there is no significant improvement over the best fit of the $\Lambda\mathrm{CDM}$ model are not shown.}
\end{figure*}

Based on the findings of the previous study \cite{Achucarro:2013cva,Achucarro:2014msa} with Planck temperature-temperature power spectrum, we sample the same region of the parameter space using \emph{only} the galaxy power spectrum from the WiggleZ Dark Energy Survey. 
As an example, Fig.\ \ref{Fig:15hr} shows residuals in the convolved galaxy power spectra, eq.\ \ref{convg}, of the feature model with parameter values $B=-0.045$, $\ln(-\tau_0)=5.55$, $\ln(\beta)=6.3$, against the base $\Lambda$CDM model in four redshift bins, with data points coming from the 15-hr region of the WiggleZ survey. The full set of the power spectra residual from all the seven fields can be found at \url{http://wwwhome.lorentz.leidenuniv.nl/~hu/links/wigglez_res/}.
%\textcolor{red}{As an example, Fig.\ \ref{Fig:15hr} shows the galaxy power spectra residual from the base-$\Lambda$CDM model in four redshift bins. The data points come from the 15-hr region of the surveys. The curves are the differences in the convolved galaxy spectra, Eq. \ref{convg}, between the best fit featured, ($B=-0.045$, $\log(-\tau_0)=5.55$, $\log(\beta)=6.3$), and base-$\Lambda$CDM models. The full set of the power spectra residual from all the seven fields can be found at \url{http://wwwhome.lorentz.leidenuniv.nl/~hu/links/wigglez_res/}.}
The parameter estimation result is shown in Fig.\ \ref{Fig:ZoomIn:WiggleZ}. In particular we show the profile likelihood of the sample in the plane $(\ln\beta,-\ln(-\tau_0))$. The upper limit of $\ln(-\tau_0)$ has been slightly extended, and the lower one slightly shrunk, in order to limit the interval to the region in which the improvement in the likelihood is significant (but we will later restore the limits of \cite{Achucarro:2013cva} in the evidence computation). 
As of the role of nuisance parameters, similarly to the results in reference \cite{Achucarro:2013cva}, no significant impact on the confidence level for the features' parameters are reported.

\begin{figure*}[htmb]
\centering
\subfigure[MCMC sampling\label{Fig:Single:mcmc}]{\includegraphics[width=0.6\textwidth]{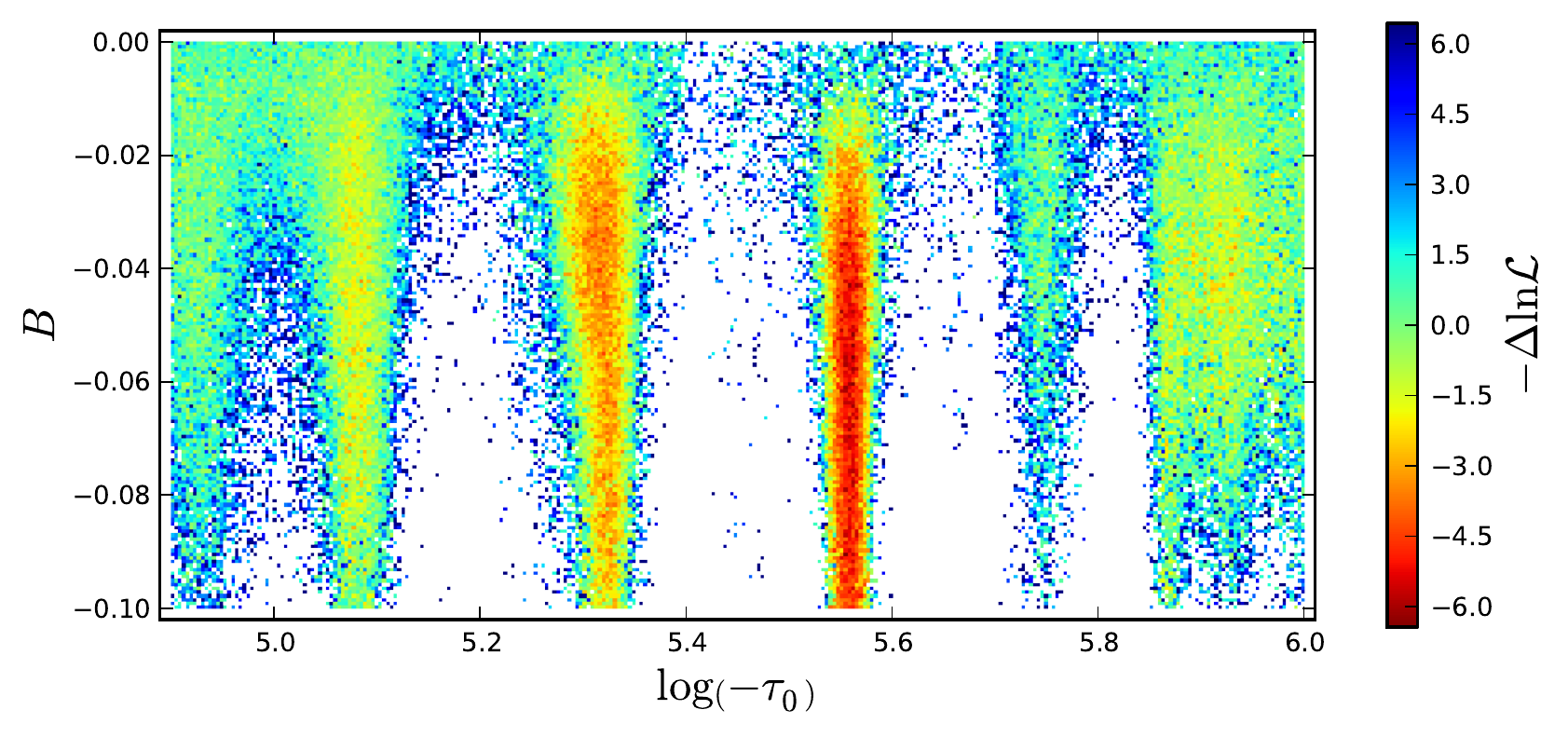}}
\subfigure[Multi-modal nested sampling\label{Fig:Single:mns}]{\includegraphics[width=0.6\textwidth]{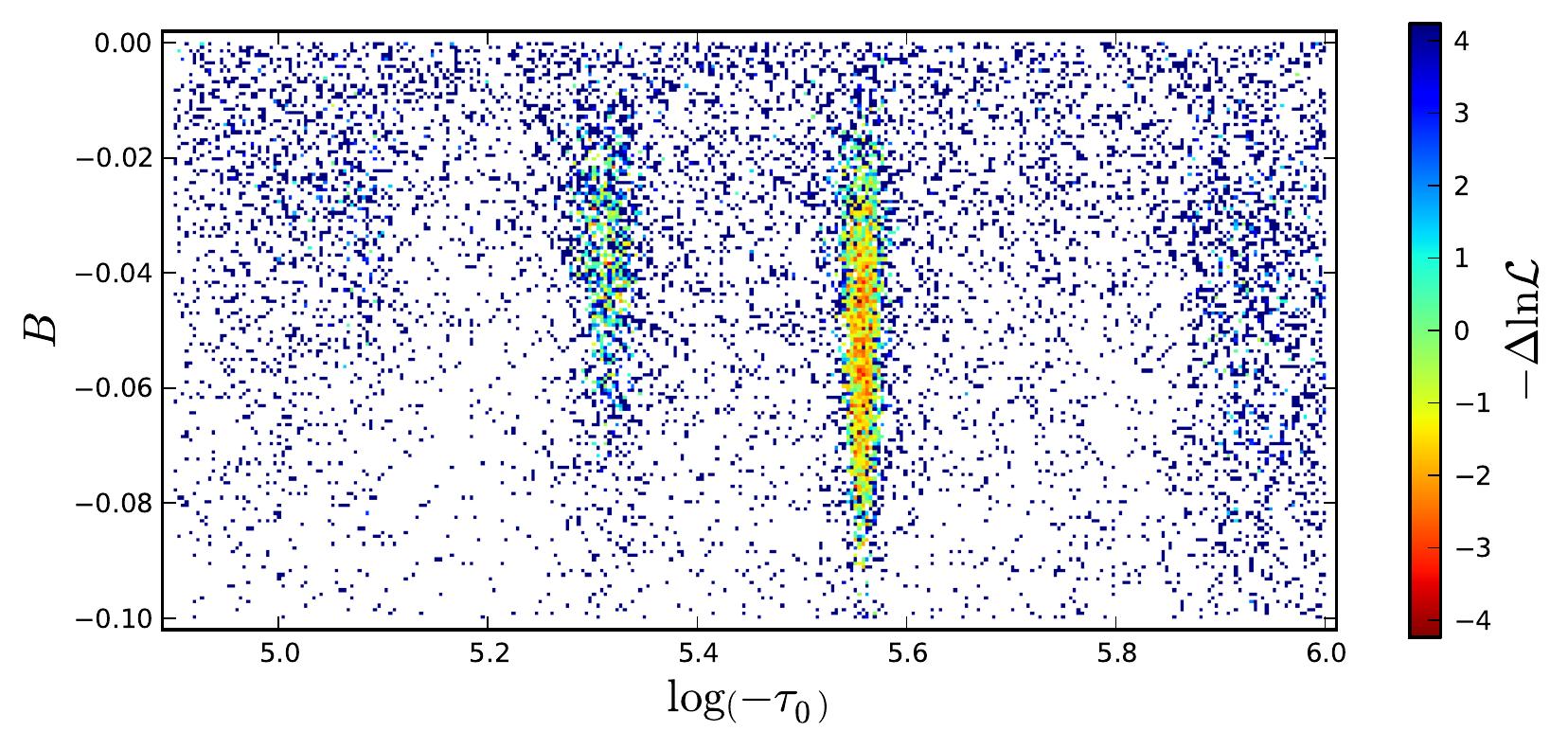}}
\caption{\label{Fig:Single}
Profile likelihood in the $(\ln(-\tau_0),B)$ plane for Planck$+$WiggleZ, for the different sampling methods. It demonstrates how the multimodal nested sampling algorithm samples more thoroughly the regions of low likelihood. The difference in the likelihood ($\Delta$) is calculated against the best fit value of $\Lambda\mathrm{CDM}$ in the different data sets.}
\end{figure*}

As we can see, in the WiggleZ posterior there exist three diffused modes. In particular, comparing Figs.\ \ref{Fig:ZoomIn:WiggleZ} and \ref{Fig:ZoomIn:Planck} with the naked eye there seems to exist a coincidence between WiggleZ and Planck results around $\ln(-\tau_0)\sim 6.0$, $\ln(-\tau_0)\sim 5.55$ and $\ln(-\tau_0)\sim 5.3$, which were three of the most significant modes detected in the previous work \cite{Achucarro:2013cva}, named respectively modes $\mathcal{A}$, $\mathcal{B}$ and $\mathcal{C}$. In order to test such coincidence, we repeated the search combining both data sets. The results are reported in Fig.~\ref{Fig:ZoomIn:PlanckWiggleZ}. The well-isolated modes previously found in the Planck data are accurately reproduced (compare Figs.\ \ref{Fig:ZoomIn:Planck} and \ref{Fig:ZoomIn:PlanckWiggleZ}, and also see Fig.\ \ref{Fig:Single:mcmc}). In addition we observe an unfolding of mode $\mathcal{A}$ and a new mode at $\ln(-\tau_0)\sim 6.3$ which survives the addition of the WiggleZ data; both of them will be the subject of future work. We have checked that there exists an enhancement of more than $20\%$  in the value of the likelihood improvement ($\Delta\ln\mathcal{L}$) in modes $\mathcal{B}$ and $\mathcal{C}$, while that of mode $\mathcal{A}$ shows no enhancement.

%%---- fig zoomin ----

At this point, let us notice two interesting characteristics of the posterior of the feature model which are present using any combination of the Planck and WiggleZ data sets. In the first place, the posterior is multi-modal in all cases. The multi-modality is due to fitting a signal with a size comparable to the noise level of the data. Second, almost every mode is elongated along the $\ln\beta$ direction, due to a degeneracy between $\ln\beta$ and $B$ that was already discussed in \cite{Achucarro:2014msa} in the context of Planck data only, and is not alleviated by including WiggleZ data.

Later, we isolated and resampled using MCMC methods each of the four individual modes found in \cite{Achucarro:2013cva} (see Fig.\ \ref{Fig:Single}) with the joint data sets. The corresponding results are shown in the Fig.~\ref{Fig:Single:mcmc}. We can see that the individual modes are separated quite well in the $\ln(-\tau_0)$ direction.

If we force ourselves to focus on one particular mode, such as mode $\mathcal{B}$, we can obtain quite stringent constraints on the feature parameters, like those demonstrated in Fig.~7 of reference \cite{Achucarro:2014msa}. However, finding stringent constraints does not mean that this result has a very strong statistical significance, because the parameter space volume of the feature model is much larger than that of the \textit{vanilla} $\Lambda\mathrm{CDM}$ model. So, even if there exists a local patch in the parameter space with highly peaked likelihood, the evidence of this signature could still be suppressed greatly by the big volume of the extra parameter space, as discussed in the previous section. 

%%---- fig Single ----

\begin{figure*}[t!]
\centering
\subfigure[1D marginalised posterior\label{Fig:multinest:1d}]{\includegraphics[width=0.8\textwidth]{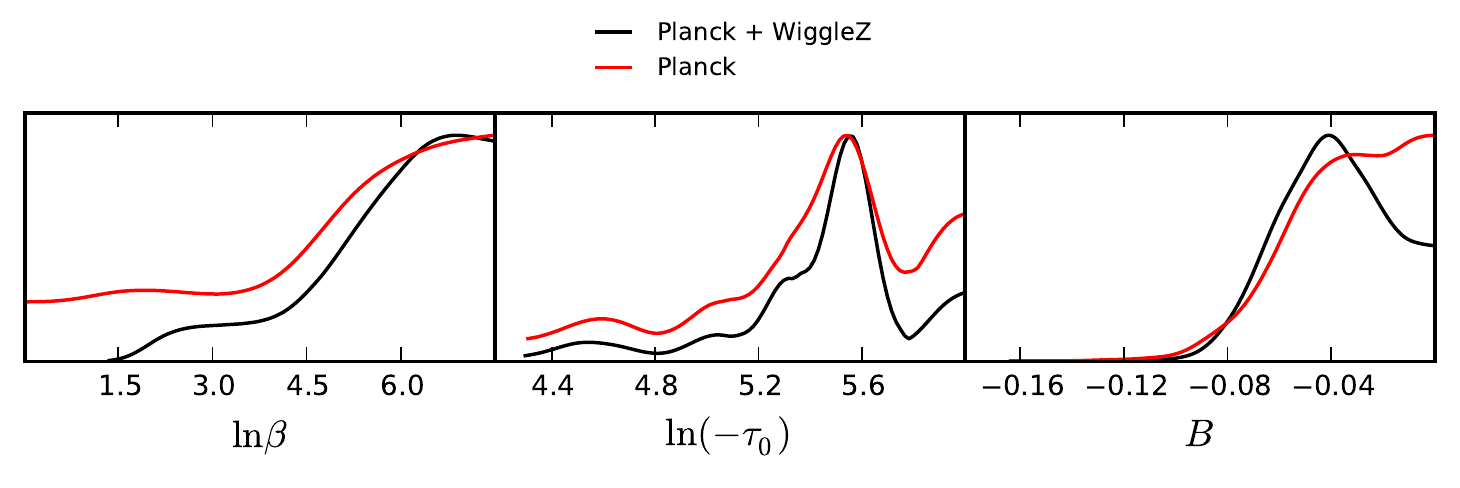}}
\subfigure[Marginalized posterior for Planck\label{Fig:multinest:2dp}]{\includegraphics[width=0.4\textwidth]{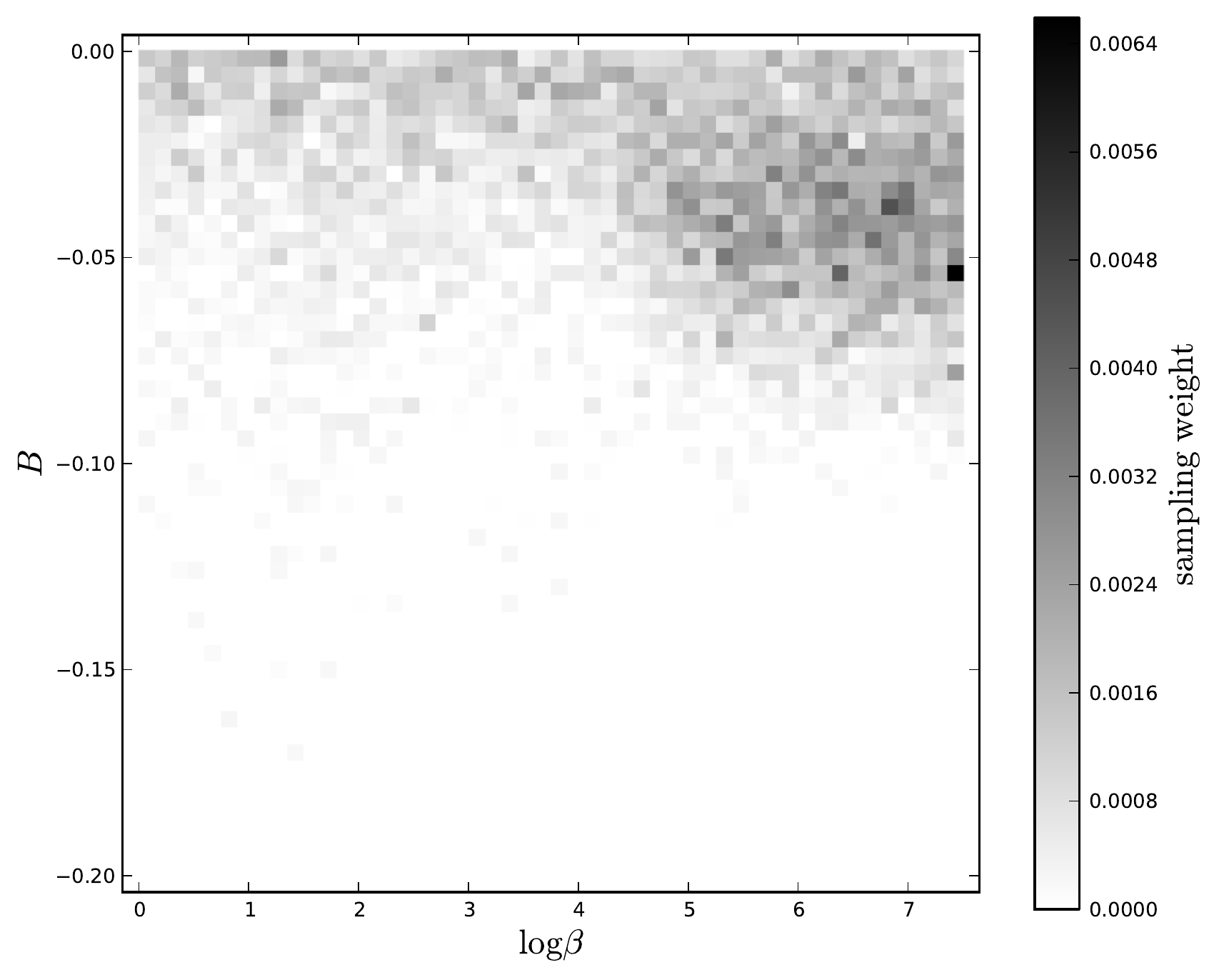}}
\subfigure[Marginalized posterior for Planck$+$WiggleZ\label{Fig:multinest:2dpw}]{\includegraphics[width=0.4\textwidth]{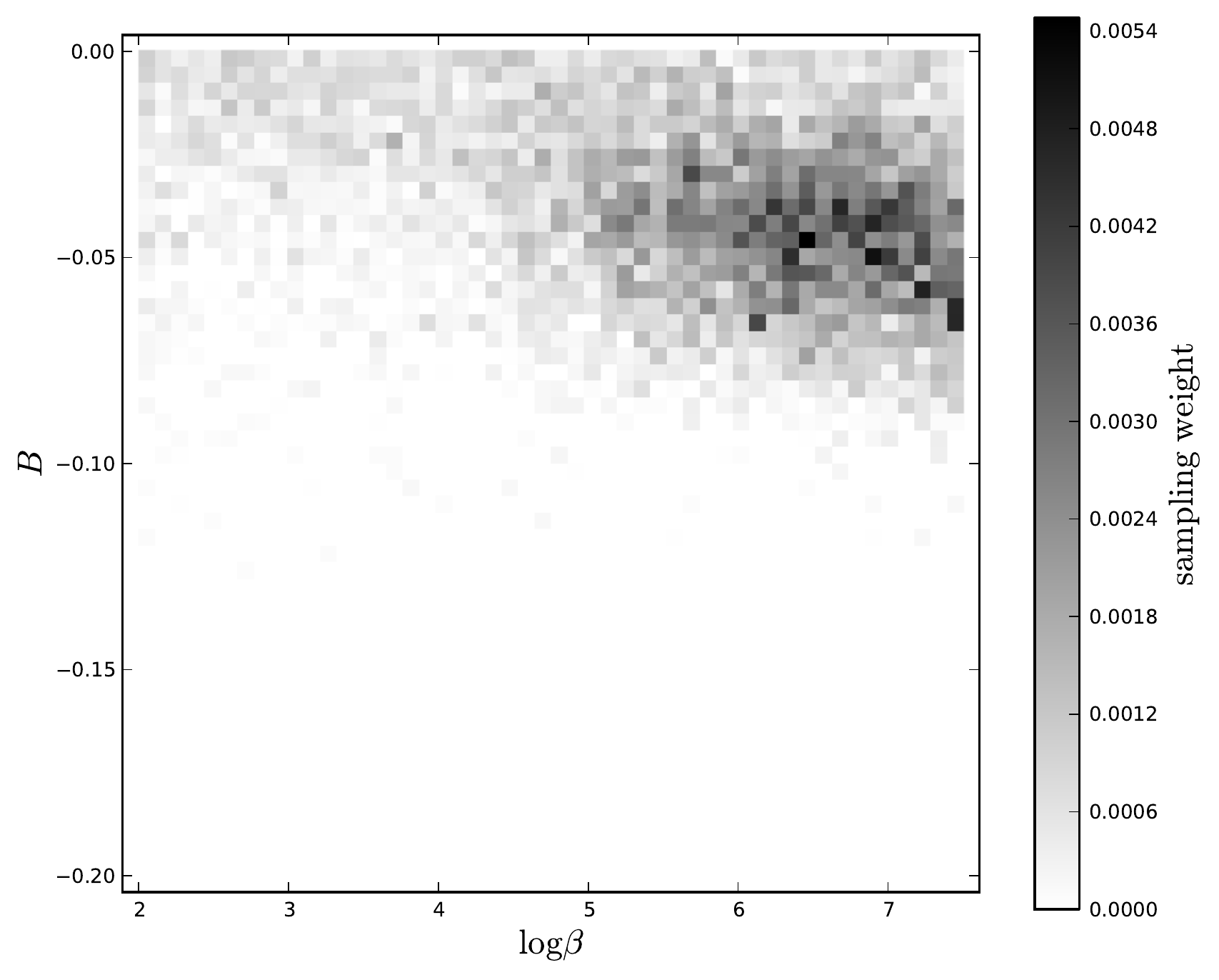}}
\caption{\label{Fig:multinest}
Multimodal nested sampling results: 1D marginalized posterior distribution for the feature parameters and 2D marginalized posterior distribution in the plane $(\ln\beta,B)$, with and without the WiggleZ data set. Notice how the addition of the WiggleZ data set increases the overall likelihood of a feature with a non-zero amplitude.}
\end{figure*}

Inspired by the fact that there exists a relatively significant reduction in the likelihood value of the feature model in the best fits compared with that of the featureless $\Lambda\mathrm{CDM}$ model, (e.g.\ for mode $\mathcal{B}$ the joint data analysis gives $-2\Delta\ln\mathcal{L}\sim10$), we are motivated to compute the Bayesian ratio of the feature model.  The statistical results are summarized in Tab.~\ref{Tab:evidence}, Fig.~\ref{Fig:Single:mns} and Fig.~\ref{Fig:multinest}. 

A comparison between the results of the MCMC and multimodal nested samplings, showing the consistency between them, can be seen in Fig.\ \ref{Fig:Single}. The main difference between both subplots is due to the more thorough sampling of the tails of the distribution (points in parameter space with low likelihood value) achieved by multi-modal nested sampling: these points are crucial to get a reliable evidence estimation, which is the goal of the nested-sampling algorithms, but almost irrelevant to parameter estimation, at which MCMC excel. In Tab.\ \ref{Tab:evidence}, we can see that the resulting best fit likelihood values from multimodal nested sampling are also consistent with those coming from MCMC sampling, though, as expected, the former a little bit lower than the latter, since the sampling around the maxima is more thorough in MCMC's.

In the first place, the Bayesian ratios listed in Tab.~\ref{Tab:evidence} tell us that there exists apparently a slightly positive preference for the feature model: $R\sim 1.9 \, (\mathrm{Planck+WiggleZ})$ vs.\ $\, 1.6 \, (\mathrm{Planck})$. However, according to the conventional criterion \cite{Jeffreys} it means that the preference is {\it barely worth mentioning}. We must emphasize that in this paper we did not cover all the parameter regime allowed by theory, which sets no lower bound for $\tau_0$, but instead the regime in which the features are most likely to be detectable by Planck. Despite the expected corrections, the slightly favourable value of the Bayesian evidence in the observable regime makes us optimistic about the enlargement of the parameter space and the addition of new data sets, namely Planck's polarization power spectrum and bispectrum. This optimistism is also backed up by how, as discussed in the Sec.~\ref{method}, the increase in the Bayesian ratio when adding the WiggleZ data indicates a positive correlation between the features found in both data sets; nevertheless, when put into the context of the error bars for the evidences cited in table \ref{Tab:evidence}, the claim gets milder.

Also, on the positive side, as can be seen in Fig.~\ref{Fig:multinest}, the addition of the WiggleZ data set clearly pushes the marginalised distribution towards bigger amplitudes of the feature with respect to using Planck data only, which on the one hand is an indication of a positive correlation between the sets, and on the other hand reinforces the overall likelihood of the presence of a feature against the null hypothesis.

% *-*-*-*-*-*-*-*-*-*-*-*-*-*-*-*
\setlength\tabcolsep{1pt}
\begin{table*}[t!]
\footnotesize
\centering
\begin{tabular}{|c|l|c|c|c|}
\hline
Model & Data set              & $-2\ln\mathcal{L}$      & $\ln\mathcal{Z}$ & $R$ \\
\hline
$M_1$ & Planck           &  $9801.918$  ($9796.27$) & $-4955.61\pm0.31$
      & \multirow{2}{*}{$\exp(0.46)\simeq 1.6$}\\
$M_0$ & Planck           &  $9807.154$  ($9805.90$) & $-4956.07\pm0.31$ &\\
\hline
$M_1$ & Planck$+$WiggleZ & $10253.570$ ($10249.20$) & $-5183.05\pm0.32$
      & \multirow{2}{*}{$\exp(0.62)\simeq 1.9$}\\
$M_0$ & Planck$+$WiggleZ & $10262.042$ ($10258.80$) & $-5183.67\pm0.31$ &\\
\hline
\end{tabular}
\caption{\label{Tab:evidence}
Multimodal nested sampling results of feature ($M_1$) and non-feature ($M_0$) models with the different data sets. The likelihood values in the third column are given at the best fit, first the nested sampling value, and second, in parenthesis, the MCMC sampling value.}
\end{table*}

%%---- fig multinest ----

%% cccccccccccccccccccccccccccccccccccccccccccccc
\section{\label{conclusion}Conclusions and outlook}
In this paper we searched for primordial oscillatory signals inspired by a transient reduction in the sound speed of the adiabatic curvature perturbation via CMB (Planck) and LSS (WiggleZ) windows. First of all, by analysing both data sets separately, we found some common oscillatory patterns both in the Planck CMB temperature-temperature power spectrum and the WiggleZ galaxy spectrum. Interestingly, we found a coincidence in the most significant mode previously found by Ach{\'u}carro \textit{et al.}\ 2013 \cite{Achucarro:2013cva} by using only Planck data. Second, the joint data analysis showed that the oscillation frequency of the feature gets better constrained, and the amplitude marginally deviates from zero, unlike what was observed by using only Planck data. Besides parameter estimation, we also calculated the Bayesian evidence for the purpose of model selection by using multimodal nested sampling. The results suggest that there exist a slightly positive preference for the feature model, Bayesian factor $R\sim 1.9~({\rm Planck+WiggleZ}),$ vs.\ $1.6~({\rm Planck})$. However, according to Jeffreys's criterion it means \textit{barely worth mentioning}.

Despite the coincidence between both data sets, we notice that the addition of LSS data does not lead to significantly higher predictivity. In theory, the galaxy surveys, compared with the CMB measurements, should be more robust at constraining features existing on scales between tens and hundreds of \unit{Mpc}, due to the fact that matter perturbations are not damped inside the sound horizon at the epoch of recombination, while they are in the photon temperature anisotropies. Nevertheless, with the present sky coverage of LSS surveys, the sample variance still dominates the uncertainties on the large scales. In order to reach a sensitivity level similar to Planck's, we need full-sky coverage and deep redshift galaxy surveys, such as {\it Euclid}.

The Bayesian evidence analysis shows that although there exists a relatively large improvement in the likelihood value ($-2\Delta\ln\mathcal L\sim10$) in several particular parameter regimes, due to the relatively large number of extra parameters ($3$) and their broad ranges of variation (look-elsewhere effect), the present Planck temperature-temperature and WiggleZ matter power spectra data still lack significance to claim a detection. However, due to the correlations between temperature and polarization modes of the power spectrum and the correlations with the bispectra given by the model of transient reductions in the speed of sound, the present results have specific \emph{predictions} for the TE cross-correlation spectrum ($C^{\rm TE}_{\ell}$) \cite{Achucarro:2014msa} and the temperature bispectrum ($B^{\rm TTT}_{\ell_1\ell_2\ell_3}$) \cite{Achucarro:2013cva,Achucarro:2014msa}. Particularly, the new fast bispectrum estimator of oscillatory features from \cite{Munchmeyer:2014nqa} should be able to cover the frequency where the most significant mode that we found is located. If those predictions are right, these signals in polarisation spectra and temperature bispectrum may be observed with the upcoming Planck data release in 2015, though it is by now unclear whether the sensitivity level of the Planck full-mission data will be high enough to claim a detection.

In the light of the additional WiggleZ data, we update the predictions stated in \cite{Achucarro:2013cva,Achucarro:2014msa}, based on the high correlation between the bispectrum features studied there and the phenomenological oscillatory shape tested by the Planck collaboration and given in \cite[eq.\ (16)]{Ade:2013ydc}.\footnote{For every combination of the feature parameter values in the regions of high likelihood, one can find a combination of the parameters in \cite{Ade:2013ydc} (including a gaussian envelope as described there) such that the correlation between both shapes at the primordial level is al least $95\%$.} In the parameters used by the Planck collaboration, we expect to find a feature with zero phase, and wavelength in the $95\%$ c.l.\ interval $k_c\in(0.0078,\,0.0083)$ from mode $\mathcal{B}$, \emph{or} $k_c\in(0.0099,\,0.0110)$ from mode $\mathcal{C}$. As happened when using only Planck data \cite{Achucarro:2014msa}, a degeneracy between $B$ and $\ln\beta$ prevents us from setting accurate predictions for the amplitude and envelope of the feature. Nevertheless, for all values of the parameters along the degeneracy, the signal is most significant on the scales beyond the second acoustic peak, and reaches its maximum around the third or fourth peak.

\begin{acknowledgments}
We are indebted to Ana Ach{\'u}carro, Enrico Pajer, Pablo Ortiz and Vicente Atal for helpful discussion. 
BH thanks Alan Heavens for helpful correspondence. This work is supported by the Dutch Foundation for Fundamental Research on Matter (FOM).
\end{acknowledgments}

%% cccccccccccccccccccccccccccccccccccccccccccccc
%\appendix


\vspace*{0.2cm}
\begin{thebibliography}{99}
  
%\cite{Bennett:2012zja}
\bibitem{Bennett:2012zja} 
  C.~L.~Bennett {\it et al.}  [WMAP Collaboration],
  %``Nine-Year Wilkinson Microwave Anisotropy Probe (WMAP) Observations: Final Maps and Results,''
  Astrophys.\ J.\ Suppl.\  {\bf 208}, 20 (2013)
  [arXiv:1212.5225 [astro-ph.CO]].
  %%CITATION = ARXIV:1212.5225;%%
  %495 citations counted in INSPIRE as of 27 Sep 2014
  
%\cite{Ade:2013uln}
\bibitem{Ade:2013uln} 
  P.~A.~R.~Ade {\it et al.}  [Planck Collaboration],
  %``Planck 2013 results. XXII. Constraints on inflation,''
  Astron.\ Astrophys.\ \textbf{571}, A22 (2014)
  [arXiv:1303.5082 [astro-ph.CO]].
  %%CITATION = ARXIV:1303.5082;%%
  %785 citations counted in INSPIRE as of 27 Sep 2014  
  
%\cite{Ade:2013ydc}
\bibitem{Ade:2013ydc} 
  P.~A.~R.~Ade {\it et al.}  [Planck Collaboration],
  %``Planck 2013 Results. XXIV. Constraints on primordial non-Gaussianity,''
  Astron.\ Astrophys.\ \textbf{571}, A24 (2014)
  [arXiv:1303.5084 [astro-ph.CO]].
  %%CITATION = ARXIV:1303.5084;%%
  %375 citations counted in INSPIRE as of 27 Sep 2014  
  
%\cite{Achucarro:2012fd}
\bibitem{Achucarro:2012fd} 
  A.~Achucarro, J.~-O.~Gong, G.~A.~Palma and S.~P.~Patil,
  %``Correlating features in the primordial spectra,''
  Phys.\ Rev.\ D {\bf 87}, 121301 (2013)
  [arXiv:1211.5619 [astro-ph.CO]].
  %%CITATION = ARXIV:1211.5619;%%
  %17 citations counted in INSPIRE as of 05 Mar 2014
  
 %\cite{Achucarro:2013cva}
\bibitem{Achucarro:2013cva} 
  A.~Achucarro, V.~Atal, P.~Ortiz and J.~Torrado,
  %``Localized correlated features in the CMB power spectrum and primordial bispectrum from a transient reduction in the speed of sound,''
  Phys.\ Rev.\ D {\bf 89}, 103006 (2014)
  [arXiv:1311.2552 [astro-ph.CO]].
  %%CITATION = ARXIV:1311.2552;%%
  %17 citations counted in INSPIRE as of 12 Sep 2014
  
%\cite{Achucarro:2014msa}
\bibitem{Achucarro:2014msa} 
  A.~Achucarro, V.~Atal, B.~Hu, P.~Ortiz and J.~Torrado,
  %``Inflation with moderately sharp features in the speed of sound: GSR and in-in formalism for power spectrum and bispectrum,''
  Phys.\ Rev.\ D {\bf 90}, 023511 (2014)
  [arXiv:1404.7522 [astro-ph.CO]].
  %%CITATION = ARXIV:1404.7522;%%
  %6 citations counted in INSPIRE as of 24 Sep 2014   
  
%\cite{Starobinsky:1992ts}
\bibitem{Starobinsky:1992ts} 
  A.~A.~Starobinsky,
  %``Spectrum of adiabatic perturbations in the universe when there are singularities in the inflation potential,''
  JETP Lett.\  {\bf 55}, 489 (1992)
  [Pisma Zh.\ Eksp.\ Teor.\ Fiz.\  {\bf 55}, 477 (1992)].
  %%CITATION = JTPLA,55,489;%%
  %200 citations counted in INSPIRE as of 27 Sep 2014
  
 %\cite{Adams:2001vc}
\bibitem{Adams:2001vc} 
  J.~A.~Adams, B.~Cresswell and R.~Easther,
  %``Inflationary perturbations from a potential with a step,''
  Phys.\ Rev.\ D {\bf 64}, 123514 (2001)
  [astro-ph/0102236].
  %%CITATION = ASTRO-PH/0102236;%%
  %160 citations counted in INSPIRE as of 27 Sep 2014
  
 %\cite{Gong:2005jr}
\bibitem{Gong:2005jr} 
  J.~O.~Gong,
  %``Breaking scale invariance from a singular inflaton potential,''
  JCAP {\bf 0507}, 015 (2005)
  [astro-ph/0504383].
  %%CITATION = ASTRO-PH/0504383;%%
  %42 citations counted in INSPIRE as of 27 Sep 2014
  
 %\cite{Chen:2008wn}
\bibitem{Chen:2008wn} 
  X.~Chen, R.~Easther and E.~A.~Lim,
  %``Generation and Characterization of Large Non-Gaussianities in Single Field Inflation,''
  JCAP {\bf 0804}, 010 (2008)
  [arXiv:0801.3295 [astro-ph]].
  %%CITATION = ARXIV:0801.3295;%%
  %180 citations counted in INSPIRE as of 27 Sep 2014
  
%\cite{Arroja:2011yu}
\bibitem{Arroja:2011yu} 
  F.~Arroja, A.~E.~Romano and M.~Sasaki,
  %``Large and strong scale dependent bispectrum in single field inflation from a sharp feature in the mass,''
  Phys.\ Rev.\ D {\bf 84}, 123503 (2011)
  [arXiv:1106.5384 [astro-ph.CO]].
  %%CITATION = ARXIV:1106.5384;%%
  %33 citations counted in INSPIRE as of 27 Sep 2014
  
%\cite{Martin:2011sn}
\bibitem{Martin:2011sn} 
  J.~Martin and L.~Sriramkumar,
  %``The scalar bi-spectrum in the Starobinsky model: The equilateral case,''
  JCAP {\bf 1201}, 008 (2012)
  [arXiv:1109.5838 [astro-ph.CO]].
  %%CITATION = ARXIV:1109.5838;%%
  %29 citations counted in INSPIRE as of 27 Sep 2014
  
%\cite{Adshead:2011jq}
\bibitem{Adshead:2011jq} 
  P.~Adshead, C.~Dvorkin, W.~Hu and E.~A.~Lim,
  %``Non-Gaussianity from Step Features in the Inflationary Potential,''
  Phys.\ Rev.\ D {\bf 85}, 023531 (2012)
  [arXiv:1110.3050 [astro-ph.CO]].
  %%CITATION = ARXIV:1110.3050;%%
  %44 citations counted in INSPIRE as of 27 Sep 2014
  
 %\cite{Arroja:2012ae}
\bibitem{Arroja:2012ae} 
  F.~Arroja and M.~Sasaki,
  %``Strong scale dependent bispectrum in the Starobinsky model of inflation,''
  JCAP {\bf 1208}, 012 (2012)
  [arXiv:1204.6489 [astro-ph.CO]].
  %%CITATION = ARXIV:1204.6489;%%
  %23 citations counted in INSPIRE as of 27 Sep 2014 
  
 %\cite{Bartolo:2013exa}
\bibitem{Bartolo:2013exa} 
  N.~Bartolo, D.~Cannone and S.~Matarrese,
  %``The Effective Field Theory of Inflation Models with Sharp Features,''
  JCAP {\bf 1310}, 038 (2013)
  [arXiv:1307.3483 [astro-ph.CO]].
  %%CITATION = ARXIV:1307.3483;%%
  %9 citations counted in INSPIRE as of 27 Sep 2014
  
  %\cite{Cannone:2014qna}
\bibitem{Cannone:2014qna} 
  D.~Cannone, N.~Bartolo and S.~Matarrese,
  %``Perturbative Unitarity of Inflationary Models with Features,''
  Phys.\ Rev.\ D {\bf 89}, 127301 (2014)
  [arXiv:1402.2258 [astro-ph.CO]].
  %%CITATION = ARXIV:1402.2258;%%
  %2 citations counted in INSPIRE as of 27 Sep 2014      
  
  %\cite{Covi:2006ci}
\bibitem{Covi:2006ci} 
  L.~Covi, J.~Hamann, A.~Melchiorri, A.~Slosar and I.~Sorbera,
  %``Inflation and WMAP three year data: Features have a Future!,''
  Phys.\ Rev.\ D {\bf 74}, 083509 (2006)
  [astro-ph/0606452].
  %%CITATION = ASTRO-PH/0606452;%%
  %104 citations counted in INSPIRE as of 29 Sep 2014
  
 %\cite{Benetti:2011rp}
\bibitem{Benetti:2011rp} 
  M.~Benetti, M.~Lattanzi, E.~Calabrese and A.~Melchiorri,
  %``Features in the primordial spectrum: new constraints from WMAP7+ACT data and prospects for Planck,''
  Phys.\ Rev.\ D {\bf 84}, 063509 (2011)
  [arXiv:1107.4992 [astro-ph.CO]].
  %%CITATION = ARXIV:1107.4992;%%
  %12 citations counted in INSPIRE as of 29 Sep 2014
  
  %\cite{Benetti:2013cja}
\bibitem{Benetti:2013cja} 
  M.~Benetti,
  %``Updating constraints on inflationary features in the primordial power spectrum with the Planck data,''
  Phys.\ Rev.\ D {\bf 88}, 087302 (2013)
  [arXiv:1308.6406 [astro-ph.CO]].
  %%CITATION = ARXIV:1308.6406;%%
  %10 citations counted in INSPIRE as of 29 Sep 2014
  
  %\cite{Hamann:2007pa}
\bibitem{Hamann:2007pa} 
  J.~Hamann, L.~Covi, A.~Melchiorri and A.~Slosar,
  %``New Constraints on Oscillations in the Primordial Spectrum of Inflationary Perturbations,''
  Phys.\ Rev.\ D {\bf 76}, 023503 (2007)
  [astro-ph/0701380].
  %%CITATION = ASTRO-PH/0701380;%%
  %82 citations counted in INSPIRE as of 29 Sep 2014
  
  %\cite{Benetti:2012wu}
\bibitem{Benetti:2012wu} 
  M.~Benetti, S.~Pandolfi, M.~Lattanzi, M.~Martinelli and A.~Melchiorri,
  %``Featuring the primordial power spectrum: new constraints on interrupted slow-roll from CMB and LRG data,''
  Phys.\ Rev.\ D {\bf 87}, 023519 (2013)
  [arXiv:1210.3562 [astro-ph.CO]].
  %%CITATION = ARXIV:1210.3562;%%
  %8 citations counted in INSPIRE as of 29 Sep 2014 
 
 %\cite{Stewart:2001cd}
\bibitem{Stewart:2001cd} 
  E.~D.~Stewart,
  %``The Spectrum of density perturbations produced during inflation to leading order in a general slow roll approximation,''
  Phys.\ Rev.\ D {\bf 65}, 103508 (2002)
  [astro-ph/0110322].
  %%CITATION = ASTRO-PH/0110322;%%
  %72 citations counted in INSPIRE as of 29 Sep 2014
  
  %\cite{Choe:2004zg}
\bibitem{Choe:2004zg} 
  J.~Choe, J.~O.~Gong and E.~D.~Stewart,
  %``Second order general slow-roll power spectrum,''
  JCAP {\bf 0407}, 012 (2004)
  [hep-ph/0405155].
  %%CITATION = HEP-PH/0405155;%%
  %61 citations counted in INSPIRE as of 29 Sep 2014
  
  %\cite{Dvorkin:2009ne}
\bibitem{Dvorkin:2009ne} 
  C.~Dvorkin and W.~Hu,
  %``Generalized Slow Roll for Large Power Spectrum Features,''
  Phys.\ Rev.\ D {\bf 81}, 023518 (2010)
  [arXiv:0910.2237 [astro-ph.CO]].
  %%CITATION = ARXIV:0910.2237;%%
  %36 citations counted in INSPIRE as of 29 Sep 2014
  
  %\cite{Adshead:2011bw}
\bibitem{Adshead:2011bw} 
  P.~Adshead, W.~Hu, C.~Dvorkin and H.~V.~Peiris,
  %``Fast Computation of Bispectrum Features with Generalized Slow Roll,''
  Phys.\ Rev.\ D {\bf 84}, 043519 (2011)
  [arXiv:1102.3435 [astro-ph.CO]].
  %%CITATION = ARXIV:1102.3435;%%
  %32 citations counted in INSPIRE as of 29 Sep 2014
  
  %\cite{Miranda:2012rm}
\bibitem{Miranda:2012rm} 
  V.~Miranda, W.~Hu and P.~Adshead,
  %``Warp Features in DBI Inflation,''
  Phys.\ Rev.\ D {\bf 86}, 063529 (2012)
  [arXiv:1207.2186 [astro-ph.CO]].
  %%CITATION = ARXIV:1207.2186;%%
  %18 citations counted in INSPIRE as of 29 Sep 2014
  
  %\cite{Adshead:2013zfa}
\bibitem{Adshead:2013zfa} 
  P.~Adshead, W.~Hu and V.~Miranda,
  %``Bispectrum in Single-Field Inflation Beyond Slow-Roll,''
  Phys.\ Rev.\ D {\bf 88}, no. 2, 023507 (2013)
  [arXiv:1303.7004 [astro-ph.CO]].
  %%CITATION = ARXIV:1303.7004;%%
  %16 citations counted in INSPIRE as of 29 Sep 2014 
  
 %\cite{Danielsson:2002kx}
\bibitem{Danielsson:2002kx} 
  U.~H.~Danielsson,
  %``A Note on inflation and transPlanckian physics,''
  Phys.\ Rev.\ D {\bf 66}, 023511 (2002)
  [hep-th/0203198].
  %%CITATION = HEP-TH/0203198;%%
  %247 citations counted in INSPIRE as of 29 Sep 2014
  
  %\cite{Greene:2004np}
\bibitem{Greene:2004np} 
  B.~R.~Greene, K.~Schalm, G.~Shiu and J.~P.~van der Schaar,
  %``Decoupling in an expanding universe: Backreaction barely constrains short distance effects in the CMB,''
  JCAP {\bf 0502}, 001 (2005)
  [hep-th/0411217].
  %%CITATION = HEP-TH/0411217;%%
  %58 citations counted in INSPIRE as of 29 Sep 2014
  
  %\cite{Meerburg:2009ys}
\bibitem{Meerburg:2009ys} 
  P.~D.~Meerburg, J.~P.~van der Schaar and P.~S.~Corasaniti,
  %``Signatures of Initial State Modifications on Bispectrum Statistics,''
  JCAP {\bf 0905}, 018 (2009)
  [arXiv:0901.4044 [hep-th]].
  %%CITATION = ARXIV:0901.4044;%%
  %126 citations counted in INSPIRE as of 29 Sep 2014
  
  %\cite{Jackson:2010cw}
\bibitem{Jackson:2010cw} 
  M.~G.~Jackson and K.~Schalm,
  %``Model Independent Signatures of New Physics in the Inflationary Power Spectrum,''
  Phys.\ Rev.\ Lett.\  {\bf 108}, 111301 (2012)
  [arXiv:1007.0185 [hep-th]].
  %%CITATION = ARXIV:1007.0185;%%
  %35 citations counted in INSPIRE as of 29 Sep 2014
  
  %\cite{Gao:2012uq}
\bibitem{Gao:2012uq} 
  X.~Gao, D.~Langlois and S.~Mizuno,
  %``Influence of heavy modes on perturbations in multiple field inflation,''
  JCAP {\bf 1210}, 040 (2012)
  [arXiv:1205.5275 [hep-th]].
  %%CITATION = ARXIV:1205.5275;%%
  %33 citations counted in INSPIRE as of 29 Sep 2014
  
  %\cite{Gao:2013ota}
\bibitem{Gao:2013ota} 
  X.~Gao, D.~Langlois and S.~Mizuno,
  %``Oscillatory features in the curvature power spectrum after a sudden turn of the inflationary trajectory,''
  JCAP \textbf{1310}, 023 (2013)
  [arXiv:1306.5680 [hep-th]].
  %%CITATION = ARXIV:1306.5680;%%
  %14 citations counted in INSPIRE as of 29 Sep 2014
  
  %\cite{Saito:2013aqa}
\bibitem{Saito:2013aqa} 
  R.~Saito and Y.~i.~Takamizu,
  %``Localized Features in Non-Gaussianity from Heavy Physics,''
  JCAP {\bf 1306}, 031 (2013)
  [arXiv:1303.3839, arXiv:1303.3839 [astro-ph.CO]].
  %%CITATION = ARXIV:1303.3839,;%%
  %12 citations counted in INSPIRE as of 29 Sep 2014
  
  %\cite{Noumi:2013cfa}
\bibitem{Noumi:2013cfa} 
  T.~Noumi and M.~Yamaguchi,
  %``Primordial spectra from sudden turning trajectory,''
  JCAP {\bf 1312}, 038 (2013)
  [arXiv:1307.7110 [hep-th]].
  %%CITATION = ARXIV:1307.7110;%%
  %9 citations counted in INSPIRE as of 29 Sep 2014 
  
  %\cite{Chen:2014joa}
\bibitem{Chen:2014joa} 
  X.~Chen and M.~H.~Namjoo,
  %``Standard Clock in Primordial Density Perturbations and Cosmic Microwave Background,''
  Phys.\ Lett.\ B \textbf{739}, 285-292 (2014)
  [arXiv:1404.1536 [astro-ph.CO]].
  %%CITATION = ARXIV:1404.1536;%%
  %2 citations counted in INSPIRE as of 25 Sep 2014
 
%\cite{Chen:2012ja}
\bibitem{Chen:2012ja} 
  X.~Chen and C.~Ringeval,
  %``Searching for Standard Clocks in the Primordial Universe,''
  JCAP {\bf 1208}, 014 (2012)
  [arXiv:1205.6085 [astro-ph.CO]].
  %%CITATION = ARXIV:1205.6085;%%
  %13 citations counted in INSPIRE as of 25 Sep 2014  
       
  %\cite{Martin:2003sg}
\bibitem{Martin:2003sg} 
  J.~Martin and C.~Ringeval,
  %``Superimposed oscillations in the WMAP data?,''
  Phys.\ Rev.\ D {\bf 69}, 083515 (2004)
  [astro-ph/0310382].
  %%CITATION = ASTRO-PH/0310382;%%
  %131 citations counted in INSPIRE as of 29 Sep 2014
  
  %\cite{Flauger:2009ab}
\bibitem{Flauger:2009ab} 
  R.~Flauger, L.~McAllister, E.~Pajer, A.~Westphal and G.~Xu,
  %``Oscillations in the CMB from Axion Monodromy Inflation,''
  JCAP {\bf 1006}, 009 (2010)
  [arXiv:0907.2916 [hep-th]].
  %%CITATION = ARXIV:0907.2916;%%
  %144 citations counted in INSPIRE as of 29 Sep 2014
  
%\cite{Flauger:2010ja}
\bibitem{Flauger:2010ja} 
  R.~Flauger and E.~Pajer,
  %``Resonant Non-Gaussianity,''
  JCAP {\bf 1101}, 017 (2011)
  [arXiv:1002.0833 [hep-th]].
  %%CITATION = ARXIV:1002.0833;%%
  %94 citations counted in INSPIRE as of 12 Oct 2014
  
  %\cite{Aich:2011qv}
\bibitem{Aich:2011qv} 
  M.~Aich, D.~K.~Hazra, L.~Sriramkumar and T.~Souradeep,
  %``Oscillations in the inflaton potential: Complete numerical treatment and comparison with the recent and forthcoming CMB datasets,''
  Phys.\ Rev.\ D {\bf 87}, 083526 (2013)
  [arXiv:1106.2798 [astro-ph.CO]].
  %%CITATION = ARXIV:1106.2798;%%
  %28 citations counted in INSPIRE as of 29 Sep 2014
  
  %\cite{Meerburg:2011gd}
\bibitem{Meerburg:2011gd} 
  P.~D.~Meerburg, R.~Wijers and J.~P.~van der Schaar,
  %``WMAP 7 Constraints on Oscillations in the Primordial Power Spectrum,''
  Mon.\ Not.\ Roy.\ Astron.\ Soc.\ {\bf 421} 369-380 (2012)
  [arXiv:1109.5264 [astro-ph.CO]].
  %%CITATION = ARXIV:1109.5264;%%
  %25 citations counted in INSPIRE as of 29 Sep 2014
  
  %\cite{Peiris:2013opa}
\bibitem{Peiris:2013opa} 
  H.~Peiris, R.~Easther and R.~Flauger,
  %``Constraining Monodromy Inflation,''
  JCAP {\bf 1309}, 018 (2013)
  [arXiv:1303.2616 [astro-ph.CO]].
  %%CITATION = ARXIV:1303.2616;%%
  %29 citations counted in INSPIRE as of 29 Sep 2014
  
  %\cite{Meerburg:2013cla}
\bibitem{Meerburg:2013cla} 
  P.~D.~Meerburg, D.~N.~Spergel and B.~D.~Wandelt,
  %``Searching for Oscillations in the Primordial Power Spectrum: Perturbative Approach (Paper I),''
  Phys.\ Rev.\ D {\bf 89}, 063536 (2014)
  [arXiv:1308.3704 [astro-ph.CO]].
  %%CITATION = ARXIV:1308.3704;%%
  %14 citations counted in INSPIRE as of 29 Sep 2014
  
 %\cite{Meerburg:2013dla}
\bibitem{Meerburg:2013dla} 
  P.~D.~Meerburg and D.~N.~Spergel,
  %``Searching for Oscillations in the Primordial Power Spectrum: Constraints from Planck (Paper II),''
  Phys.\ Rev.\ D {\bf 89}, 063537 (2014)
  [arXiv:1308.3705 [astro-ph.CO]].
  %%CITATION = ARXIV:1308.3705;%%
  %16 citations counted in INSPIRE as of 29 Sep 2014 
      
 %\cite{Achucarro:2010da}
\bibitem{Achucarro:2010da} 
  A.~Achucarro, J.~O.~Gong, S.~Hardeman, G.~A.~Palma and S.~P.~Patil,
  %``Features of heavy physics in the CMB power spectrum,''
  JCAP {\bf 1101}, 030 (2011)
  [arXiv:1010.3693 [hep-ph]].
  %%CITATION = ARXIV:1010.3693;%%
  %110 citations counted in INSPIRE as of 29 Sep 2014
  
 %\cite{Hu:2011vr}
\bibitem{Hu:2011vr} 
  W.~Hu,
  %``Generalized Slow Roll for Non-Canonical Kinetic Terms,''
  Phys.\ Rev.\ D {\bf 84}, 027303 (2011)
  [arXiv:1104.4500 [astro-ph.CO]].
  %%CITATION = ARXIV:1104.4500;%%
  %26 citations counted in INSPIRE as of 29 Sep 2014
  
 %\cite{Park:2012rh}
\bibitem{Park:2012rh} 
  M.~Park and L.~Sorbo,
  %``Sudden variations in the speed of sound during inflation: features in the power spectrum and bispectrum,''
  Phys.\ Rev.\ D {\bf 85}, 083520 (2012)
  [arXiv:1201.2903 [astro-ph.CO]].
  %%CITATION = ARXIV:1201.2903;%%
  %23 citations counted in INSPIRE as of 29 Sep 2014
  
 %\cite{Nakashima:2010sa}
\bibitem{Nakashima:2010sa} 
  M.~Nakashima, R.~Saito, Y.~i.~Takamizu and J.~Yokoyama,
  %``The effect of varying sound velocity on primordial curvature perturbations,''
  Prog.\ Theor.\ Phys.\  {\bf 125}, 1035 (2011)
  [arXiv:1009.4394 [astro-ph.CO]].
  %%CITATION = ARXIV:1009.4394;%%
  %24 citations counted in INSPIRE as of 29 Sep 2014
  
 %\cite{Bean:2008na}
\bibitem{Bean:2008na} 
  R.~Bean, X.~Chen, G.~Hailu, S.-H.~H.~Tye and J.~Xu,
  %``Duality Cascade in Brane Inflation,''
  JCAP {\bf 0803}, 026 (2008)
  [arXiv:0802.0491 [hep-th]].
  %%CITATION = ARXIV:0802.0491;%%
  %51 citations counted in INSPIRE as of 29 Sep 2014
  
 %\cite{Ribeiro:2012ar}
\bibitem{Ribeiro:2012ar} 
  R.~H.~Ribeiro,
  %``Inflationary signatures of single-field models beyond slow-roll,''
  JCAP {\bf 1205}, 037 (2012)
  [arXiv:1202.4453 [astro-ph.CO]].
  %%CITATION = ARXIV:1202.4453;%%
  %22 citations counted in INSPIRE as of 29 Sep 2014      
  
  %\cite{Meerburg:2010ks}
\bibitem{Meerburg:2010ks} 
  P.~D.~Meerburg,
  %``Oscillations in the Primordial Bispectrum I: Mode Expansion,''
  Phys.\ Rev.\ D {\bf 82}, 063517 (2010)
  [arXiv:1006.2771 [astro-ph.CO]].
  %%CITATION = ARXIV:1006.2771;%%
  %20 citations counted in INSPIRE as of 06 Oct 2014
  
  %\cite{Meerburg:2014kna}
\bibitem{Meerburg:2014kna} 
  P.~D.~Meerburg, D.~N.~Spergel and B.~D.~Wandelt,
  %``Searching for oscillations in the primordial power spectrum,''
  arXiv:1406.0548 [astro-ph.CO].
  %%CITATION = ARXIV:1406.0548;%%
  %1 citations counted in INSPIRE as of 06 Oct 2014

  %\cite{Achucarro:2010jv}
\bibitem{Achucarro:2010jv} 
  A.~Achucarro, J.~O.~Gong, S.~Hardeman, G.~A.~Palma and S.~P.~Patil,
  %``Mass hierarchies and non-decoupling in multi-scalar field dynamics,''
  Phys.\ Rev.\ D {\bf 84}, 043502 (2011)
  [arXiv:1005.3848 [hep-th]].
  %%CITATION = ARXIV:1005.3848;%%
  %57 citations counted in INSPIRE as of 29 Sep 2014
  
 %\cite{Cespedes:2012hu}
\bibitem{Cespedes:2012hu} 
  S.~Cespedes, V.~Atal and G.~A.~Palma,
  %``On the importance of heavy fields during inflation,''
  JCAP {\bf 1205}, 008 (2012)
  [arXiv:1201.4848 [hep-th]].
  %%CITATION = ARXIV:1201.4848;%%
  %47 citations counted in INSPIRE as of 29 Sep 2014
  
  %\cite{Achucarro:2012sm}
\bibitem{Achucarro:2012sm} 
  A.~Achucarro, J.~O.~Gong, S.~Hardeman, G.~A.~Palma and S.~P.~Patil,
  %``Effective theories of single field inflation when heavy fields matter,''
  JHEP {\bf 1205}, 066 (2012)
  [arXiv:1201.6342 [hep-th]].
  %%CITATION = ARXIV:1201.6342;%%
  %56 citations counted in INSPIRE as of 29 Sep 2014 
  
   %\cite{Lewis:1999bs}
\bibitem{Lewis:1999bs} 
  A.~Lewis, A.~Challinor and A.~Lasenby,
  %``Efficient computation of CMB anisotropies in closed FRW models,''
  Astrophys.\ J.\  {\bf 538}, 473 (2000),
  [astro-ph/9911177].
  %%CITATION = ASTRO-PH/9911177;%%
  %1277 citations counted in INSPIRE as of 06 Dec 2013
   
 %\cite{Lewis:2002ah}
\bibitem{Lewis:2002ah} 
  A.~Lewis and S.~Bridle,
  %``Cosmological parameters from CMB and other data: A Monte Carlo approach,''
  Phys.\ Rev.\ D {\bf 66}, 103511 (2002)
  [astro-ph/0205436].
  %%CITATION = ASTRO-PH/0205436;%%
  %1428 citations counted in INSPIRE as of 12 Sep 2014     
  
 %\cite{Feroz:2008xx}
\bibitem{Feroz:2008xx} 
  F.~Feroz, M.~P.~Hobson and M.~Bridges,
  %``MultiNest: an efficient and robust Bayesian inference tool for cosmology and particle physics,''
  Mon.\ Not.\ Roy.\ Astron.\ Soc.\  {\bf 398}, 1601 (2009)
  [arXiv:0809.3437 [astro-ph]].
  %%CITATION = ARXIV:0809.3437;%%
  %297 citations counted in INSPIRE as of 10 Sep 2014

%\cite{Feroz:2007kg}
\bibitem{Feroz:2007kg} 
  F.~Feroz and M.~P.~Hobson,
  %``Multimodal nested sampling: an efficient and robust alternative to MCMC methods for astronomical data analysis,''
  Mon.\ Not.\ Roy.\ Astron.\ Soc.\  {\bf 384}, 449 (2008)
  [arXiv:0704.3704 [astro-ph]].
  %%CITATION = ARXIV:0704.3704;%%
  %228 citations counted in INSPIRE as of 10 Sep 2014
  
\bibitem{2013arXiv1306.2144F} 
Feroz F., Hobson M.~P., Cameron E., Pettitt A.~N., 2013, arXiv, 
[arXiv:1306.2144 [astro-ph.IM]].
  

%\cite{Ade:2013zuv}
\bibitem{Ade:2013zuv} 
  P.~A.~R.~Ade {\it et al.}  [Planck Collaboration],
  %``Planck 2013 results. XVI. Cosmological parameters,''
  Astron.\ Astrophys.\ \textbf{571}, A16 (2014)
  [arXiv:1303.5076 [astro-ph.CO]].
  %%CITATION = ARXIV:1303.5076;%%
  %1354 citations counted in INSPIRE as of 04 Mar 2014

%\cite{Ade:2013kta}
\bibitem{Ade:2013kta} 
  P.~A.~R.~Ade {\it et al.}  [Planck Collaboration],
  %``Planck 2013 results. XV. CMB power spectra and likelihood,''
  Astron.\ Astrophys.\ \textbf{571}, A15 (2014)
  [arXiv:1303.5075 [astro-ph.CO]].
  %%CITATION = ARXIV:1303.5075;%%
  %149 citations counted in INSPIRE as of 04 Mar 2014
  
 %\cite{Parkinson:2012vd}
\bibitem{Parkinson:2012vd} 
  D.~Parkinson, S.~Riemer-Sorensen, C.~Blake, G.~B.~Poole, T.~M.~Davis, S.~Brough, M.~Colless and C.~Contreras {\it et al.},
  %``The WiggleZ Dark Energy Survey: Final data release and cosmological results,''
  Phys.\ Rev.\ D {\bf 86}, 103518 (2012)
  [arXiv:1210.2130 [astro-ph.CO]].
  %%CITATION = ARXIV:1210.2130;%%
  %33 citations counted in INSPIRE as of 04 Mar 2014 
  
%\cite{Drinkwater:2009sd}
\bibitem{Drinkwater:2009sd} 
  M.~J.~Drinkwater, R.~J.~Jurek, C.~Blake, D.~Woods, K.~A.~Pimbblet, K.~Glazebrook, R.~Sharp and M.~B.~Pracy {\it et al.},
  %``The WiggleZ Dark Energy Survey: Survey Design and First Data Release,''
  Mon.\ Not.\ Roy.\ Astron.\ Soc.\  {\bf 401}, 1429 (2010)
  [arXiv:0911.4246 [astro-ph.CO]].
  %%CITATION = ARXIV:0911.4246;%%
  %138 citations counted in INSPIRE as of 04 Mar 2014   

%\cite{Blake:2010xz}
\bibitem{Blake:2010xz} 
  C.~Blake, S.~Brough, M.~Colless, W.~Couch, S.~Croom, T.~Davis, M.~J.~Drinkwater and K.~Forster {\it et al.},
  %``The WiggleZ Dark Energy Survey: the selection function and z=0.6 galaxy power spectrum,''
  Mon.\ Not.\ Roy.\ Astron.\ Soc.\  {\bf 406}, 803 (2010)
  [arXiv:1003.5721 [astro-ph.CO]].
  %%CITATION = ARXIV:1003.5721;%%
  %44 citations counted in INSPIRE as of 04 Mar 2014

%\cite{Riemer-Sorensen:2013jsa}
\bibitem{Riemer-Sorensen:2013jsa} 
  S.~Riemer-Sørensen, D.~Parkinson and T.~M.~Davis,
  %``Combining Planck with Large Scale Structure gives strong neutrino mass constraint,''
  Phys.\ Rev.\ D \textbf{89} 103505 (2014)
  [arXiv:1306.4153 [astro-ph.CO]].
  %%CITATION = ARXIV:1306.4153;%%
  %16 citations counted in INSPIRE as of 04 Mar 2014
  
\bibitem{Skilling}
Skilling J., 2004, AIP Conference Proceedings of the 24th International
Workshop on Bayesian Inference and Maximum Entropy
Methods in Science and Engineering, Vol. 735, pp. 395-405  

\bibitem{Sivia}
Sivia D., Skilling J., 2006, Data Analysis; a Bayesian tutorial,
2nd ed., Oxford University Press, Oxford
  
%\cite{Mukherjee:2005wg}
\bibitem{Mukherjee:2005wg}
  P.~Mukherjee, D.~Parkinson and A.~R.~Liddle,
  %``A nested sampling algorithm for cosmological model selection,''
  Astrophys.\ J.\  {\bf 638} (2006) L51
  [astro-ph/0508461].
  %%CITATION = ASTRO-PH/0508461;%%
  %94 citations counted in INSPIRE as of 25 Sep 2014  
  
%\cite{Liddle:2007fy}
\bibitem{Liddle:2007fy} 
  A.~R.~Liddle,
  %``Information criteria for astrophysical model selection,''
  Mon.\ Not.\ Roy.\ Astron.\ Soc.\  {\bf 377}, L74 (2007)
  [astro-ph/0701113].
  %%CITATION = ASTRO-PH/0701113;%%
  %120 citations counted in INSPIRE as of 25 Sep 2014
  
 %\cite{Shaw:2007jj}
\bibitem{Shaw:2007jj} 
  R.~Shaw, M.~Bridges and M.~P.~Hobson,
  %``Clustered nested sampling: Efficient Bayesian inference for cosmology,''
  Mon.\ Not.\ Roy.\ Astron.\ Soc.\  {\bf 378}, 1365 (2007)
  [astro-ph/0701867 [ASTRO-PH]].
  %%CITATION = ASTRO-PH/0701867;%%
  %21 citations counted in INSPIRE as of 12 Oct 2014 
  
\bibitem{Jeffreys}
S. H. Jeffreys, {\it The Theory of Probability} (Oxford University Press, New York, NY, USA, 1961).    
    
%\cite{Munchmeyer:2014nqa}
\bibitem{Munchmeyer:2014nqa} 
  M.~M\"unchmeyer, F.~Bouchet, M.~Jackson and B.~Wandelt,
  %``The Komatsu Spergel Wandelt estimator for oscillations in the cosmic microwave background bispectrum,''
  Astron.\ Astrophys.\ \textbf{570} A94 (2014)
  [arXiv:1405.2550 [astro-ph.CO]].
  %%CITATION = ARXIV:1405.2550;%%
  %1 citations counted in INSPIRE as of 06 Oct 2014
  

  

      
  
    
  
  
\end{thebibliography}
\end{document}